\documentclass[11pt,a4paper]{article}
\pdfoutput=1

\usepackage{jheppub}
 
\usepackage{graphicx, epsfig} 
\usepackage{subfig}
\usepackage{amsmath,amssymb,amsfonts,dsfont,mathrsfs,amsthm,mathtools}
\usepackage{bm} 
\usepackage{color}
\usepackage[usenames]{xcolor}
\usepackage{hyperref}
\usepackage{siunitx}
\hypersetup{colorlinks=true,urlcolor=blue,linkcolor=magenta,citecolor=blue,filecolor=blue}
\usepackage[normalem]{ulem}
\usepackage{array}
\usepackage{hyperref}
\usepackage{booktabs}
\usepackage{blindtext}
\usepackage{upgreek}
\usepackage{subfloat}
\usepackage[normalem]{ulem}
\newcommand{\stkout}[1]{\ifmmode\text{\sout{\ensuremath{#1}}}\else\sout{#1}\fi}

\allowdisplaybreaks

\usepackage{float}
\usepackage[scr=rsfs]{mathalfa}
\usepackage{bigints}
\usepackage[english]{babel}
\usepackage{float}
\usepackage{braket}

\newcommand{\diff}[1]{\text{d}#1}

\begin{document}

\title{Renormalization of Einstein-Gauss-Bonnet AdS gravity}

\author[a]{Giorgos Anastasiou,}
\author[b]{Ignacio J. Araya,}
\author[b]{Avik Chakraborty,}
\author[c]{Crist\'obal Corral,}
\author[d]{Rodrigo Olea}

\affiliation[a]{Departamento de Ciencias, Facultad de Artes Liberales, Universidad Adolfo Ibáñez, \\ Avenida Diagonal Las Torres 2640, 7941169, Pe\~nalol\'en, Chile \vspace{0.1cm}}
\affiliation[b]{Departamento de F\'isica y Astronom\'ia, Facultad de Ciencias Exactas, Universidad Andres Bello,  \\ Sazi\'e 2212, Piso 7, Santiago, Chile \vspace{0.1cm} \vspace{0.1cm}}
\affiliation[c]{Departamento de Ciencias, Facultad de Artes Liberales, Universidad Adolfo Ibáñez, \\ Avenida Padre Hurtado 750, 2562340, Viña del Mar, Chile \vspace{0.1cm}}
\affiliation[d]{Instituto de Física, Pontificia Universidad Católica de Valparaíso, Casilla 4059, Valparaíso, Chile \vspace{0.1cm}}

\vspace{0.3cm}
\emailAdd{georgios.anastasiou@uai.cl} 
\emailAdd{ignacio.araya@unab.cl}
\emailAdd{avik.phys88@gmail.com}
\emailAdd{cristobal.corral@uai.cl}
\emailAdd{rodrigo\_olea\_a@yahoo.co.uk}

\abstract{
The asymptotic analysis for the metric of a generic solution of Einstein-Gauss-Bonnet AdS theory is provided by solving the field equations in the Fefferman-Graham frame. Using standard holographic renormalization, the counterterms that render the action finite are found up to seven spacetime dimensions. In the case of 6D, an equivalent formulation that permits a fully covariant determination of the counterterms is introduced, based on the finiteness of conformal invariants. It is shown that both schemes end up in the same holographic stress-energy tensor. Physical properties of six-dimensional topological Boulware-Deser black holes in  Einstein-Gauss-Bonnet-AdS$_6$ gravity, whose boundary has nontrivial conformal features, are worked out in detail. Employing both renormalization prescriptions,  finite asymptotic charges are found, and the correct black hole thermodynamics is recovered.}

\maketitle
\section{Introduction}

Higher-curvature corrections in gravity theories appear in different contexts of theoretical physics. Their presence is inevitable if gravity is regarded as an effective theory from a Wilsonian viewpoint. The Gauss-Bonnet term, for instance, arises from the low-energy limit of string theory~\cite{Zwiebach1985, Zwiebach2004}. Higher-curvature corrections are also induced by quantum anomalies~\cite{Graham:1999pm, AlvarezGaume:1983ig, Capper:1975ig, Deser:1993yx, Fernandes:1983ts} and, in the case of quadratic gravity, the theory becomes one-loop renormalizable in Minkowski space, with the price of losing unitarity due to the presence of ghosts~\cite{Stelle:1976gc}. In the Anti-de Sitter/Conformal Field Theory (AdS/CFT) correspondence~\cite{Maldacena:1997re, Aharony:1999ti}, on the other hand, higher-curvature terms induce novel holographic sources at the conformal boundary, allowing one to compute quantum responses for the boundary theory~\cite{Skenderis:2002wp}. Also, the inclusion in the gravity action of terms of higher order in the curvature describes more generic CFT duals with different central charges, beyond what can be obtained in Einstein gravity~\cite{Henningson:1998gx}. However, the renormalization of infrared divergencies in asymptotically AdS spacetimes turns more challenging when higher-curvature terms are present. The resolution of this problem has not been achieved for generic gravitational solutions.

Among the infinite family of theories with higher order in the curvature, Lovelock gravity~\cite{Lovelock:1971yv} can be regarded as the natural generalization of Einstein theory in higher dimensions. Its higher-curvature terms correspond to dimensionally-continued Euler topological densities and their field equations remain second-order, propagating the same number of degrees of freedom~\cite{henneaux1990}. The next term in the Lovelock series after the Ricci scalar is the Gauss-Bonnet density, which becomes dynamical in dimensions higher than four, generating interesting solutions such as black holes~\cite{boulware1985,wheeler1986} and gravitational instantons~\cite{Dehghani:2006aa,Corral:2019leh,Corral:2022udb,Corral:2025yvr}, even with boundaries which are not conformally flat. On the other hand, the Gauss-Bonnet term contributes nontrivially to the black hole entropy, modifying the standard area law of Einstein gravity~\cite{Jacobson:1993xs, Wald:1993nt,Cai:2001dz}.

In the context of AdS/CFT, it has been shown that particular values of the Gauss-Bonnet coupling might violate the shear viscosity to entropy bound in the dual theory; a feature which has been conjectured to be universal in any holographic dual to Einstein-AdS gravity~\cite{Brigante:2007nu}. This provides a holographic argument for constraining the Gauss-Bonnet coupling such that the dual CFT is consistent. Additionally, in Refs.~\cite{deBoer:2009pn,Buchel:2009sk} the relation between the Gauss-Bonnet coupling and the universal coefficients arising in correlators of the holographic stress tensor in the dual CFT was studied. This implies a correction to the central charges, modifying the coefficients of the conformal anomaly in the dual CFT. Furthermore, it can be used to set causality and positivity bounds on the propagation of fluctuations as shown in Refs.~\cite{deBoer:2009pn,Buchel:2009sk}.\footnote{Holographic aspects when the Gauss-Bonnet term is nonminimally coupled to scalar fields have been explored in Refs.~\cite{Caceres:2023gfa,Tong:2025rxz}.}

Holographically, a renormalized gravitational on-shell action is needed for a well-defined generating functional of the dual CFT theory, in agreement with the GKPW formula~\cite{Gubser:1998bc, Witten:1998qj}. This requires the action to have a well-posed variational principle with respect to the holographic sources. This is also the case of Lovelock gravity in its linearized form.

For asymptotically AdS (AAdS) spacetimes, both the sources and conjugate operators can be identified geometrically as certain terms in the Fefferman-Graham (FG) form of the metric,
\begin{equation}\label{gaussnormal}
    \diff{s^2} = \frac{\ell^2}{z^2}\left(\diff{z^2} + \bar{g}_{ij}(x,z)\diff{x^i}\diff{x^j}\right)\,.
\end{equation}
Here, $z$ denotes the holographic Poincaré coordinate,  $\{x^i\}$ are the coordinates at the conformal boundary, where the corresponding metric can be expanded as
\begin{equation}
     \bar{g}_{ij}(x,z) = g_{(0)ij} + \frac{z^2 }{\ell^2}g_{(2)ij} + \frac{z^4}{\ell^4}g_{(4)ij} + \ldots \,,
     \label{FG-exp}
\end{equation}
where, in Einstein gravity in $d+1$ dimensions, only even powers of $z$ appear until the holographic order $\mathcal{O}(z^d)$. The conformal boundary is located at $z=0$ while the metric $g_{(0)ij}$ is identified with the holographic source of the CFT stress-tensor. Pure Einstein gravity ---as well as Lovelock gravity in its linearized form having the same degrees of freedom--- requires a variational problem with respect to $g_{(0)ij}$ ---i.e, the CFT background metric.

Motivated by the locality of the fields in the Wilsonian interpretation of QFT, the authors in Ref.~\cite{Emparan:1999pm, hep-th/9910267, Balasubramanian:1999re, deHaro:2000vlm, Henningson:1998gx, Skenderis:2002wp}, introduced the holographic renormalization prescription, which moreover ensures a Dirichlet variational principle with respect to the induced metric at finite radius $h_{ij}$. As a consequence, one has to add the Gibbons-Hawking-York (GHY) term in the case of Einstein gravity or the Myers terms~\cite{Myers1987} for Lovelock gravity and then introduce a series of boundary covariant counterterms which cancel all the infrared divergences in a minimal subtraction scheme (i.e., without modifying the universal part).

In the method outlined above, the equations of motion in the radial decomposition fix the subleading coefficients in the expansion \eqref{FG-exp} in terms of the holographic data $g_{(0)ij}$. In practice, however, finding the higher-order coefficients in terms of the $g_{(0)ij}$ using the field equations is technically hard for arbitrary higher-derivative gravity theories. Furthermore, the boundary terms required for fixing the Dirichlet variational principle with respect to the induced metric at finite radius are only known in the case of Lovelock gravity (Myers terms in Ref.~\cite{Myers1987}). For these reasons, the Holographic Renormalization (HR) counterterms are known for generic $g_{(0)ij}$ for Einstein-AdS in a low enough dimension. In turn, in Lovelock gravity, they are only known for a conformally flat boundary $g_{(0)ij}$ metric~\cite{Yale:2011dq} by assuming the form for the lowest-derivative counterterms, as covariant functionals of the boundary metric.

For Einstein-AdS gravity in even bulk dimensions, a covariant renormalization scheme exists, which makes use of symmetry principles to ensure both the finiteness and the Dirichlet variational principle with respect to the $g_{(0)ij}$ source. This scheme is called Conformal Renormalization (CR)~\cite{Anastasiou:2016jix, Anastasiou:2020mik, Anastasiou:2022ljq, Anastasiou:2023oro}, and it considers embedding the theory into Conformal Gravity. Upon imposing the Einstein condition on the spacetime, the theory reduces to Einstein gravity with the necessary counterterms to achieve the renormalization. The method has the advantage that the counterterms are given directly from the embedding and are fixed. The justification for the prescription relies on the fact that all Einstein-AdS spacetimes are solutions of Conformal Gravity (CG) in four dimensions. In six dimensions, one can construct a specific combination of conformal invariants where this is still true. Then, as all Einstein solutions are also solutions of a given CG theory, one can choose this particular sector to evaluate the action. Then, the finiteness of Conformal Gravity for AAdS spaces gets inherited by the resulting corrected Einstein gravity action. The obtained action has a specific form, which considers the Einstein-Hilbert (EH) term, the topological Euler density with a specific coupling, and an extra boundary term that cancels remaining divergences that depend on the boundary Weyl tensor. The EH plus Euler part corresponds to the Topological Renormalization action~\cite{Aros:1999id,Olea:2006vd, Miskovic:2009bm, 0708.0782, Miskovic:2014zja, Anastasiou:2020zwc}, and the extra boundary term can be thought of as a correction, although it comes naturally from the CR prescription and therefore it is fixed by conformal symmetry. It is important to stress that the additional term is a fully covariant contribution proportional to $\Box \lvert F \rvert^2$, where
\begin{equation}\label{AdS_curvature}
    F^{\mu \nu}_{\lambda \rho}=R^{\mu \nu}_{\lambda \rho}+\frac{1}{\ell^2_{\rm eff}}\delta^{\mu \nu}_{\lambda \rho}
\end{equation}
is the Riemannian part of the bulk AdS curvature and can be thought of as a constant displacement of the Riemann, such that it is equal to zero for pure AdS.

In the case of Lovelock gravity, one can simply follow the procedure described above to perform the Holographic Renormalization of the theory. For an arbitrary boundary metric $g_{(0)ij}$ and for the radial decomposition of the Lovelock equations of motion, the higher-order coefficients are obtained in terms of the holographic data. When used in the bulk action plus the Myers terms, the FG frame produces a divergent power series in $z$. The divergent pieces are explicitly dependent on the FG coefficients. Therefore, the construction of counterterms as covariant quantities of the boundary metric $h_{ij}$ requires the inversion of the series. In this work, these holographic techniques are applied in full generality for the case of Einstein-Gauss-Bonnet-AdS (EGB-AdS) gravity in arbitrary dimension up to order $z^{5}$ in the Poincaré holographic coordinate.

In a different vein, it is also shown here that adding the Euler term to the even-dimensional EGB-AdS action (Topological Renormalization) does not recover the results of Holographic Renormalization. The boundary terms responsible for the mismatch, however, vanish for conformally flat boundaries. In six dimensions, this situation can be corrected by considering derivative terms in a fully covariant form, in a similar fashion as in Einstein gravity. We work out this case in full detail here.

The paper is organized as follows. Section~\ref{S2} introduces Einstein-Gauss-Bonnet theory, discussing the equations of motion and vacua of the theory. The condition for the existence of a degenerate vacuum state is also analyzed. Section~\ref{asymptoticanalysis} considers the asymptotic analysis of the generic solutions of the theory, by introducing the FG expansion in terms of the holographic Poincaré coordinate z. The equations of motion are expanded in the FG gauge, and from them, the relation of the subleading coefficients of the metric in terms of the holographic data $g_{(0)ij}$ is determined (in the non-degenerate case). From this expansion, the bulk Weyl tensor is computed to all relevant orders in z, separating it into the AdS curvature $F$ defined in Eq.~\eqref{AdS_curvature} and a tensor $X$ which is subleading when the two vacua are different. In Section~\ref{S4}, using the asymptotic form obtained for AAdS solutions to EGB, the Holographic Renormalization counterterms of the theory are worked out. Also, from the resulting renormalized EGB gravity action, the holographic stress-tensor is derived at the conformal boundary of the spacetime. Sections~\ref{Section 4} and \ref{sec:CR} present a different way of renormalizing the EGB-AdS theory in six dimensions, which gives identical results for holographic quantities at the conformal boundary but which is fully covariant in the spacetime metric. This procedure builds upon the addition of a topological term~\cite{Aros:1999id,Aros:1999kt,Olea:2005gb,Olea:2006vd,0708.0782, Kofinas:2008ub, Miskovic:2009bm, Miskovic:2014zja, Anastasiou:2020zwc}, but incorporates an extra total derivative which cancels remaining sub-leading divergences present for manifolds with non-conformally-flat boundaries. In Section~\ref{Section 4}, this total derivative is motivated by relating it to the conformal invariants in six dimensions, whereas the cancellation of the remaining divergences is verified in Section~\ref{sec:CR}.  The method has the advantage of facilitating the computation of holographic quantum information-theoretic measures~\cite{Anastasiou:2017xjr, Anastasiou:2018mfk, Anastasiou:2018rla, Anastasiou:2019ldc, Anastasiou:2021swo, Anastasiou:2021jcv}. It can be used for different metric gauges that are not of the FG form. Indeed, examples of non-Gaussian coordinates consider frames to study properties of conformal fluids in the context of the fluid/gravity correspondence~\cite{Rangamani:2009xk}. It also includes the possibility of an adapted gauge in the analysis of flat gravity, which has a well-defined asymptotically Minkowski limit~\cite{Compere:2019bua}. In Section~\ref{S6}, Boulware-Deser black holes in EGB theory with non-conformally flat transverse sections and non-constant curvature at the horizon are studied. These topological black holes are the generalization to EGB theory of the solutions for Einstein-AdS gravity analyzed in Ref.~\cite{Anastasiou:2023oro}. They serve as a test ground for computing their renormalized on-shell action using the scheme discussed in Section~\ref{sec:CR}. Indeed, in Section~\ref{S7}, the conserved charges and thermodynamical quantities are computed for the black holes found in this work. Finally, Section~\ref{S8} summarizes the main results and comments on their meaning and physical relevance.

\section{Einstein-Gauss-Bonnet AdS gravity}\label{S2}

 Einstein-Gauss-Bonnet gravity is a particular sector of the Lanczos-Lovelock theory, which considers corrections to General Relativity up to quadratic terms in the curvature in higher dimensions. Its dynamics in arbitrary $d+1$ dimensions is dictated by the action functional
\begin{align}\label{Ibulk}
    I_{\rm EGB}[g_{\mu\nu}] &= \kappa\int\limits_{\mathcal{M}}\diff{^{d+1}x}\sqrt{|g|}\left(R-2\Lambda + \alpha\mathcal{G} \right)\,,
\end{align}
where $\kappa=(16\pi G)^{-1}$ is the gravitational constant, $\Lambda$ is the cosmological constant, and $\alpha$ is a dimensionful coupling constant. The Gauss-Bonnet term is defined as
\begin{align}
    \mathcal{G} = \frac{1}{4}\delta^{\mu_1\ldots\mu_4}_{\nu_1\ldots\nu_4}R^{\nu_1\nu_2}_{\mu_1\mu_2}R^{\nu_3\nu_4}_{\mu_3\mu_4} = R^2 - 4R_{\mu\nu}R^{\mu\nu} + R_{\mu\nu\lambda\rho}R^{\mu\nu\lambda\rho}.
\end{align}
\subsection{Equations of motion}
In four dimensions, the bulk integral of the Gauss-Bonnet term can be written as the sum of the boundary integral of the Chern form plus the Euler characteristic via the Gauss-Bonnet theorem.

In higher dimensions, however, the Gauss-Bonnet term contributes to the field equations in a nontrivial way. The latter are obtained by performing arbitrary variations of the action~\eqref{Ibulk} with respect to the metric, giving 
\begin{align}\label{eom}
 \mathcal{E}_{\mu\nu} :=  R_{\mu\nu} - \frac{1}{2}g_{\mu\nu}R + \Lambda g_{\mu\nu} + \alpha H_{\mu\nu} = 0\,,
\end{align}
where $G_{\mu \nu}= R_{\mu\nu} - \tfrac{1}{2}g_{\mu\nu}R + \Lambda g_{\mu\nu}$ is the Einstein tensor in the presence of the cosmological constant and $H_{\mu\nu}$ is the Lanczos tensor, which is explicitly given by
\begin{align}\label{HGB}
 H^\mu_\nu = -\frac{1}{8}\delta^{\mu\mu_1\ldots\mu_4}_{\nu\nu_1\ldots\nu_4}R^{\nu_1\nu_2}_{\mu_1\mu_2}R^{\nu_3\nu_4}_{\mu_3\mu_4} =  2R^{\mu\rho}_{\sigma\lambda}R_{ \nu\rho}^{\sigma\lambda}-4R_\rho^\sigma R^{\mu\rho}_{ \nu\sigma}+2RR^\mu_\nu - 4R^\mu_\lambda R^\lambda_\nu - \frac{1}{2}\delta^\mu_\nu \mathcal{G}\,.
\end{align}
This particular combination of higher-curvature terms gives rise to second-order field equations for the metric. Diffeomorphism invariance implies that this term is covariantly conserved. As mentioned in the introduction, the space of solutions of the field equations~\eqref{eom} has been studied in different contexts.

The contribution of the Gauss-Bonnet term in Eq.~\eqref{eom} modifies the asymptotics of the solutions, generating an effective AdS radius $\ell_{\rm eff}$. This constant is defined implicitly in terms of the bare AdS radius 
\begin{equation}
\ell^2=-\frac{d(d-1)}{2\Lambda}\,,
\end{equation}
and the Gauss-Bonnet coupling through the relation
\begin{align}\label{leff}
 \frac{1}{\ell^2} = \frac{ 1}{\ell_{\rm eff}^2} -\frac{\alpha  \left(d-2\right) \left(d-3\right)}{\ell_{\rm eff}^4}\,.
\end{align}
This is a quadratic equation for the effective AdS radius possessing two real roots, $\ell^{+}_{\rm eff}$ and $\ell^{-}_{\rm eff}$. Each of these roots corresponds to a different vacuum of the theory, such that the field equations~\eqref{eom} can be written in factorized form as
\begin{align}\label{eom_fact}
\delta^{\mu\mu_1\ldots\mu_4}_{\nu\nu_1\ldots\nu_4}\left(R^{\nu_1\nu_2}_{\mu_1\mu_1} + \frac{1}{\left(\ell_{\rm eff}^{+}\right)^{2}}\delta^{\nu_1\nu_2}_{\mu_1\mu_2} \right)\left(R^{\nu_3\nu_4}_{\mu_3\mu_4} + \frac{1}{\left(\ell_{\rm eff}^{-}\right)^2}\delta^{\nu_3\nu_4}_{\mu_3\mu_4} \right) = 0\,.
\end{align}
Additionally, there exists a point in parameter space where the two maximally symmetric vacua coalesce. This point is characterized by the condition
\begin{align}\label{lamdadegenerated}
\Lambda=-\frac{d\left(d-1\right)}{8\alpha \left(d-2\right) \left(d-3\right)}\,, 
\end{align}
which gives a degenerate effective AdS radius with a value\footnote{In five dimensions, in the so-called Chern-Simons point, the holographic description of the theory is radically different, which is reflected in the computation of the corresponding Weyl anomaly \cite{Banados:2004zt,Banados:2005rz}.}
\begin{align}\label{degenerated}
\ell_{\text{deg}}^2 := \left(\ell_{\rm eff}^{+}\right)^2 = \left(\ell_{\rm eff}^{-}\right)^2 =\frac{\ell^2}{2}=2 \left(d-2\right) \left(d-3\right)\alpha \,. 
\end{align}
As a feature commonly present in gravity theories with AdS asymptotia, the action for EGB-AdS contains infrared divergences, which come from the double pole in $z$ of the line element at the conformal boundary. In the following sections, a detailed analysis of the asymptotic structure of the theory is provided, what is helpful to determine the counterterms that render finite both the action~\eqref{Ibulk} and the corresponding charges.

\subsection{Asymptotic analysis} \label{asymptoticanalysis}

In order to work out the counterterms for EGB in AAdS spacetimes, one considers the FG expansion defined in Eqs.~\eqref{gaussnormal} and \eqref{FG-exp}, using the effective AdS radius instead,\footnote{From a mathematical point of view, the FG expansion for Lovelock metrics was investigated in Refs.~\cite{Albin:2005qka,yu2025conformallycompactmetricslovelock}.} what amounts to the change of $\ell \rightarrow \ell_{\rm eff}$. Then, this asymptotic expansion of the metric is replaced in the equations of motion (EOM) to find the subleading coefficients in terms of the boundary data $g_{(0)ij}$. Initially, the EOM are expressed in their independent components of the radial decomposition. To do so, one needs to introduce the Gauss-Codazzi relations, given in Appendix~\ref{GC}. It is important to determine the expansion of the extrinsic curvature in this gauge
\begin{equation}
K^i_j = \frac{1}{\ell_{\rm eff}} \delta^i_j
-\frac{z^2}{\ell_{\rm eff}^3} g^i_{\left(2\right)j} + \frac{z^4}{\ell_{\rm eff}^5} \left(g^{i m}_{\left(2\right)}g_{\left(2\right)mj}- 2g^i_{\left(4\right)j}\right) + \mathcal{O} \left(z^6\right) \,.
\end{equation}
Then the asymptotic expansion of the field equations for the $zz$, $ij$ and $zi$ components reads
\begin{align}
G^{z}_{z} +\alpha H^{z}_{z} &= \frac{d \left(d-1\right) }{2}\Delta^{\left(0\right)} - \frac{z^2}{ 2 \ell_{\rm eff}^4} \Delta^{\left(1\right)} \left[ 2 \left(d-1\right)g_{\left(2\right)}+\ell_{\rm eff}^2 \mathcal{R}_{\left(0\right)}\right] + \mathcal{O}\left(z^4\right) \,, \notag \\
G^{i}_{j} +\alpha H^{i}_{j} &= \frac{d \left(d-1\right) }{2} \Delta^{\left(0\right)} \delta^{i}_{j} - \frac{z^2}{2\ell_{\rm eff}^4} \Delta^{\left(1\right)} \Big[2 \left(d-2\right) \left(g_{\left(2\right)} \delta^{i}_{j} -  g_{\left(2\right)j}^i\right) \notag\\ & - 2 \ell_{\rm eff}^2 \mathcal{R}^{i}_{\left(0\right)j} + \ell_{\rm eff}^2 \delta^{i}_{j} \mathcal{R}_{\left(0\right)}\Big] +\mathcal{O}\left(z^4\right) \,, \notag \\
G^{z}_{j} +\alpha H^{z}_{j} &= \frac{2 z^{3}}{\ell^4} \Delta^{\left(1\right)} \mathcal{D}_{\left(0\right) [m}g_{\left(2\right)j]}^{m} +\mathcal{O}\left(z^5\right) \,.
\end{align}
From now on, objects in caligraphic will denote codimension-1, while the subscript $(0)$ stands for tensors constructed out of $g_{(0)ij}$, such that $\mathcal{R}^{i}_{(0)jkl}$ is the Riemann tensor of $g_{(0)ij}$ and $\mathcal{D}_{(0)i}$ denotes the Levi-Civita covariant derivative which is metric-compatible with $g_{(0)ij}$. Additionally, $\Delta^{\left(0\right)}$ represents the characteristic polynomial that defines the effective curvature radius of the two maximally symmetric vacua. This is given by the relation
\begin{equation}
\Delta^{\left(0\right)} = -\frac{1}{\ell^2} + \frac{1}{\ell_{\rm eff}^2} - \frac{\alpha \left(d-2\right) \left(d-3\right) }{\ell_{\rm eff}^4} \,, 
\end{equation}
and vanishes identically by virtue of Eq.~\eqref{leff}. As a consequence, the leading-order term of the field equations depends only on the polynomial that defines the degeneracy condition, i.e.,
\begin{equation}
\Delta^{\left(1\right)} =\frac{d\Delta^{(0)}}{d\ell^{-2}_{\rm eff}}= 1-\frac{2 \left(d-2\right) \left(d-3\right) \alpha}{\ell_{\rm eff}^2} \,.
\end{equation}
For nondegenerate vacua, i.e., $\Delta^{\left(1\right)}\neq0$, one can solve $g_{\left(2\right)ij}$ from the field equations explicitly. In such a case, it takes the form of Poincar\'e-Einstein manifolds~\cite{Skenderis:2000in, deHaro:2000vlm}, namely,
\begin{equation}
g_{\left(2\right)ij} = - \ell_{\rm eff}^2 \mathcal{S}_{\left(0\right)ij} \,,
\label{g2schouten}
\end{equation}
where $\mathcal{S}_{\left(0\right)ij}$ is the Schouten tensor of the boundary metric $g_{\left(0\right)ij}$ defined as
\begin{equation}\label{Schouten}
\mathcal{S}_{(0)i}^{j} = \frac{1}{d-2}\left[\mathcal{R}^i_{(0)j} - \frac{1}{2(d-1)}\delta^i_j \mathcal{R}_{(0)} \right]\,.
\end{equation}
 Similarly, one determines the next term in the FG expansion, which can be cast as
\begin{align}\label{g4}
g^i_{(4)j}
&= \frac{\ell^4_{\rm eff}}{4(d-4)}\left[(d-4)\mathcal{S}^i_{(0)k}\mathcal{S}^k_{(0)j} - \mathcal{B}^i_{(0)j} - \frac{4\alpha }{\ell^2_{\rm eff}\Delta^{(1)}}\left((\mathcal{W}^2_{(0)})^i_j-\frac{3}{4(d-1)}\delta^i_j|\mathcal{W}_{(0)}|^2 \right)\right]\,,
\end{align}
where $\mathcal{W}^{i}_{(0)jkl}$ is the Weyl tensor of $g_{(0)ij}$, $(\mathcal{W}_{(0)}^2)^i_j := \mathcal{W}_{(0)}^{iklm}\mathcal{W}_{(0)jklm}$, and the Bach and Cotton tensors defined in terms of the boundary metric are 
\begin{equation}\label{Bach}
\mathcal{B}_{(0)j}^{i} = \mathcal{D}^{k}_{(0)} \mathcal{C}^{i}_{(0)jk}+\mathcal{S}_{(0)k}^{l}\mathcal{W}^{k i}_{(0
)lj}\,,
\end{equation}
and
\begin{equation}\label{Cotton}
 \mathcal{C}_{(0)ijk} = 2\mathcal{D}_{(0)[k}\mathcal{S}_{(0)j]i}\,,
\end{equation}
respectively. 

The coefficient $g_{\left(2\right)ij}$ in Eq.~\eqref{g2schouten} has a universal form  (up to the definition of $\ell_{\rm eff}$), out of the degenerate point of the theory. In a similar fashion, from Eq.~\eqref{g4}, one notices that $g_{\left(4\right) ij}$ can be decomposed into a universal part and a theory-dependent piece, with an explicit dependence on the Gauss-Bonnet coupling. The same conclusion can be drawn from a purely kinematic description, in terms to Penrose-Brown-Henneaux (PBH) transformations in the asymptotic form of the metric, mimicking the analysis in Ref.~\cite{Brown:1986nw,hep-th/9910267,Penrose:1985bww}.

Finally, the trace of the next order in the FG expansion, $g_{(6)}$, is determined, as it is relevant for the computation of the Weyl anomaly in seven dimensions. In particular, one obtains
\begin{align}
g_{(6)}&=\frac{\ell^6_{\rm eff}}{6(d-4)}\mathcal{B}^{ab}_{(0)}\mathcal{S}_{(0)ab} + \frac{\alpha \ell^4_{\rm eff}}{12(d-1)(d-4)\Delta^{(1)}}\left[2(d-4)\mathcal{D}_a(\mathcal{W}^{abcd}_{(0)}\mathcal{C}_{(0)bcd})  \right. \nonumber\\ 
&\left. +4(d-3)(d-4)|\mathcal{C}_{(0)}|^2-6\mathcal{S}_{(0)}|\mathcal{W}_{(0)}|^2 + 2 \mathcal{S}^{ab}_{(0)}\mathcal{W}^{cde}_{(0)a}[(3d-2)\mathcal{W}_{(0)bcde}+4\mathcal{W}_{(0)bdce}]  \right. \nonumber\\
& \left. -4\mathcal{W}_{(0)}^{abcd}\mathcal{W}_{(0)a c}^{e f}\mathcal{W}_{(0)bedf} - 2\mathcal{W}_{(0)}^{abcd}\mathcal{W}_{(0)ab}^{ef}\mathcal{W}_{(0)cedf} - 4\mathcal{W}_{(0)abcd}\mathcal{W}_{(0)}^{befc}\mathcal{W}^{ad}_{(0)e f} \right. \nonumber \\ 
&\left. + \mathcal{W}^{ab}_{(0)cd}\mathcal{W}^{cd}_{(0)ef}\mathcal{W}^{ef}_{(0)ab} \right] \,,
\end{align}
where $|\mathcal{C}_{(0)}|^2:=\mathcal{C}_{(0)abc} \mathcal{C}_{(0)}^{abc}$. It is important to stress that all terms in the FG expansion --including the odd-power modes-- involve the degeneracy polynomial $\Delta^{(1)}$, such that when it vanishes, the series remains undetermined. This behavior shows that the standard asymptotic analysis breaks down when a degenerate vacuum is considered. In this work, the generic case $\Delta^{\left(1\right)} \neq 0$ is studied.

There are features shared by both Einstein and EGB gravity, which become evident in the asymptotic behavior of the Weyl tensor. The latter can be decomposed on-shell as~\cite{Jatkar:2015ffa}
\begin{equation}
W^{\alpha \beta}_{\mu \nu} = R^{\alpha \beta}_{\gamma \delta}-4\delta^{[\alpha}_{[\gamma} S^{\beta]}_{\delta ]} = F^{\alpha \beta}_{\mu \nu} + X^{\alpha \beta}_{\mu \nu} \, ,
\label{Weyl-F}
\end{equation}
where $S^\mu_\nu$ is the bulk Schouten tensor, and
\begin{equation}
X^{\alpha \beta}_{\mu \nu} = \left[\frac{1}{\ell^2}-  \frac{1}{\ell_{\rm eff}^2} -\frac{2}{d \left(d-1\right)}\alpha H\right] \delta^{\alpha \beta}_{\mu \nu} +\frac{4}{\left(d-1\right)}\alpha H^{[\alpha}_{[\mu} \delta^{\beta]}_{\nu]} \,,
\end{equation}
represents the non-Einstein part of the Weyl and $H=g^{\mu\nu}H_{\mu\nu}$. This definition measures the deviation from a maximally symmetric space with curvature radius $\ell_{\rm eff}$.

The asymptotic expansion of the components of $F^{\alpha \beta}_{\mu \nu}$ are
\begin{align}
F^{i z}_{j z} &= \frac{z^4}{\ell_{\rm eff}^6} \left(g_{\left(2\right)}^{im} g_{\left(2\right) jm} - 4 g_{\left(4\right)j}^{i}\right) +\mathcal{O} \left(z^6\right) \,, \notag \\
F^{i j}_{k m} &= \frac{z^2}{\ell_{\rm eff}^2} \left( \mathcal{R}^{ij}_{\left(0\right) km} +\frac{4}{\ell_{\rm eff}^2} g_{\left(2\right) [k}^{[i} \delta^{j]}_{m]} \right) +\mathcal{O} \left(z^4\right) \,, \notag \\
F^{i z}_{j k} &= \frac{2 z^3}{\ell_{\rm eff}^4}  \mathcal{D}_{\left(0\right) [j}g_{\left(2\right)k]}^{i} + \mathcal{O} \left(z^5\right) \,.
\label{Ftensorcompon}
\end{align}
In turn, asymptotic expansion of $X^{\alpha\beta}_{\mu\nu}$ can be expressed in the FG frame as
\begin{align}
\label{Xtensorcompon}
X^{i z}_{j z} &= \frac{2 \left(d-2\right) \left(d-3\right)\alpha z^2 }{d \left(d-1\right)\ell_{\rm eff}^6} \Big[\ell_{\rm eff}^2 \mathcal{R}_{\left(0\right)} \delta^{i}_{j}  - d \ell_{\rm eff}^{2} \mathcal{R}^{i}_{\left(0\right) j} \notag \\ &\qquad \qquad - d \left(d-2\right) g_{\left(2\right) j}^{i} + \left(d-2\right) g_{\left(2\right)} \delta^{i}_{j} \Big] +\mathcal{O} \left(z^4\right) \,, \notag \\
X^{i j}_{k m} &=  \frac{2 \left(d-2\right) \left(d-3\right)  \alpha z^2}{d \left(d-1\right)  \ell_{\rm eff}^6} \Big(2 g_{\left(2\right)} \delta^{i j}_{k m} + 4 d \left(d-2\right)  g_{\left(2\right) [k}^{[i}  \delta^{j]}_{m]} \notag \\ &\qquad \qquad +4 d  \ell_{\rm eff}^{2} \mathcal{R}^{[i}_{\left(0\right) [k} \delta^{j]}_{m]} - \ell_{\rm eff}^{2}  \mathcal{R}_{\left(0\right)} \delta^{i j}_{k m} \Big) +\mathcal{O} \left(z^4\right) \,, \notag \\
X^{i z}_{j k} &= - \frac{4 \left[d \left(d-5\right) +3\right] \alpha z^{3}}{\left(d-1\right) \ell_{\rm eff}^6}  \delta^{i}_{[j} \left(\mathcal{D}_{\left(0\right) m} g_{\left(2\right)k]}^{m} - \mathcal{D}_{\left(0\right) k]} g_{\left(2\right)}\right) + \mathcal{O} \left(z^5\right) \,.
\end{align}
In the nondegenerate case, the dynamics expresses the previous formulae in terms of tensorial quantities of the boundary. For the $F$ tensor, one gets
\begin{align}
F^{i z}_{j z} &= \frac{z^4}{\left(d-4\right) \ell_{\rm eff}^2 } \left[\mathcal{B}^{i}_{\left(0\right)j} + \frac{4\alpha }{\ell^2_{\rm eff}\Delta^{(1)}}\left((\mathcal{W}^2_{(0)})^i_j-\delta^i_j|\mathcal{W}_{(0)}|^2 \right)\right] +\mathcal{O} \left(z^6\right) \,, \notag \\
F^{i j}_{k m} &= \frac{z^2}{\ell_{\rm eff}^2} \mathcal{W}^{ij}_{\left(0\right) km}  +\mathcal{O} \left(z^4\right) \,,  \notag \\
F^{i z}_{j k} &= - \frac{ z^3}{\ell_{\rm eff}^2}  \mathcal{C}_{(0)j k}^{i} + \mathcal{O} \left(z^5\right) \,.
\end{align}
In a similar way, for the $X$ tensor one obtains
\begin{align}\label{Xtensorcomponnodeg}
X^{i z}_{j z} &= \frac{\alpha z^4 }{d \left(d-1\right) \Delta^{(1)} \ell_{\rm eff}^4 }  \left(2 d (\mathcal{W}^2_{(0)})^i_j - 3\delta^i_j|\mathcal{W}_{(0)}|^2 \right)+\mathcal{O} \left(z^6\right) \,, \notag \\
X^{i j}_{k m} &=  \frac{\alpha z^4 }{d \left(d-1\right) \Delta^{(1)} \ell_{\rm eff}^4 } \left(2 d (\mathcal{W}^2_{(0)})^{[i}_{[k} \delta^{j]}_{m]} - 3|\mathcal{W}_{(0)}|^2 \delta^{i j}_{k m}\right)+\mathcal{O} \left(z^6\right)  \,, \notag \\
X^{i z}_{j k} &=  \mathcal{O} \left(z^5\right) \,.
\end{align}
Notice that the leading-order contributions in Eq.~\eqref{Xtensorcompon} vanish due to Eq.~\eqref{g2schouten}, which now fall off at least as $\mathcal{O}\left(z^4\right)$. As a consequence, the asymptotic behavior of the Weyl tensor is dominated by $F^{\mu\nu}_{\lambda\rho}$, what resembles the Einstein-AdS case where the $X$ tensor vanishes identically. In this sense, one may characterize Lanczos-Lovelock spacetimes as asymptotically Einstein. Furthermore, using the definition in Eq.~\eqref{Weyl-F} and the asymptotic expansion given in Eq.~\eqref{Xtensorcomponnodeg}, one can show that
\begin{equation}\label{S-asymp}
    S^\nu_\lambda = -\frac{1}{2\ell^2_{\rm eff}}\delta^\nu_\lambda -\frac{1}{4}X^{\mu\nu}_{\mu\lambda}+\frac{1}{40}X^{\mu\rho}_{\mu\rho}\delta^\nu_\lambda = -\frac{1}{2\ell^2_{\rm eff}}\delta^\nu_\lambda + \mathcal{O}(z^4). 
\end{equation}
Finally, the asymptotic expansion of the bulk Cotton tensor is given by 
\begin{subequations}
    \begin{align}
C^{i}_{zj} &=- \frac{z}{d \left(d-1\right) \ell^4_{\rm eff}} \left[ d\left(d-2\right)\left(g_{\left(2\right)j}^{i}-\frac{1}{d}\delta^{i}_{j} g_{\left(2\right)}  \right)+ d \ell^2_{\rm eff} \left(\mathcal{R}_{\left(0\right)j}^{i} - \frac{1}{d}\delta^{i}_{j} \mathcal{R}_{\left(0\right)}\right)\right]+ \mathcal{O} \left(z^3\right) \,,\\
\notag
C^{i}_{jk} &= -\frac{ z^2}{d\left(d-1\right)\ell^4_{\rm eff}} \left[2d\delta^{i}_{[j} \mathcal{D}_{\left(0\right)m} g_{\left(2\right)k]}^{m} - 2\left(d+1\right)\delta^{i}_{[j} \mathcal{D}_{\left(0\right)k]} g_{\left(2\right)} +2d \left(d-2\right)\mathcal{D}_{\left(0\right)[j} g_{\left(2\right)k]}^{i}  \right. \notag \\
& \left. + 2d \ell^2_{\rm eff} \mathcal{D}_{\left(0\right)[j} \mathcal{R}_{\left(0\right)k]}^{i} + \ell^2_{\rm eff} \delta^{i}_{[j} \mathcal{D}_{\left(0\right)k]} \mathcal{R}_{\left(0\right)}\right] +  \mathcal{O} \left(z^4\right) \,, \\
C^{z}_{jk} &= \frac{2 z^3}{\left(d-1\right)\ell^4_{\rm eff}} \left(\mathcal{D}_{\left(0\right)[j} \mathcal{D}_{\left(0\right)m}  g_{\left(2\right)k]}^{m} + g_{\left(2\right)[j}^{m} \mathcal{R}_{\left(0\right)k] m} \right) +  \mathcal{O} \left(z^5\right)\,, \\
C^{z}_{iz} &= \frac{z^2}{2d\left(d-1\right)\ell^4_{\rm eff}} \left[2 \left(d-2\right) \mathcal{D}_{\left(0\right)i}  g_{\left(2\right)} -2d \mathcal{D}_{\left(0\right)m}  g_{\left(2\right)i}^{m} + \ell^2_{\rm eff} \mathcal{D}_{\left(0\right)i]} \mathcal{R}_{\left(0\right)}\right] + \mathcal{O} \left(z^4 \right) \,.
\end{align}
\end{subequations}
Notice that, in the non-degenerate case, the leading orders of the Cotton vanish such that
\begin{align}
C^{i}_{zj} &=\frac{ 6 \alpha z^3}{\left(d-1\right) \Delta^{(1)} \ell^4_{\rm eff}} \left((\mathcal{W}^2_{(0)})^i_j-\frac{1}{d}\delta^i_j|\mathcal{W}_{(0)}|^2 \right)  + \mathcal{O} \left(z^5\right)  \,, \label{lqc}\\ C^{i}_{jk} &\sim \mathcal{O} \left(z^4\right) \,, \quad  
C^{z}_{jk} \sim  \mathcal{O} \left(z^5\right) \,, \quad
C^{z}_{iz} \sim  \mathcal{O} \left(z^4 \right) \,.
\end{align}
In what follows, the asymptotic expansion of the objects above will be used the renormalized action and quasilocal stress tensor.

\section{Holographic renormalization and renormalized stress tensor}\label{S4}

The previous analysis allows us to determine the structure of the infrared divergences in the EGB-AdS action in arbitrary dimensions. The appearance of the GB coupling in $g_{\left(4\right) ij}$ implies the existence of new divergences with respect to the Einstein case at fourth-derivative order. Hence, it is expected that the counterterms in EGB would not be only a modification of the corresponding couplings in the case of Einstein gravity, but an inclusion of  additional terms which account for conformal properties of the boundary. Along that line, in what follows, the boundary dynamics is worked out on a basis consisting of the Schouten and the Weyl tensors, and derivatives thereof, i.e., the Cotton and Bach tensors. Therefore, the corresponding counterterms are written in terms of these tensors, what is more natural in manifolds endowed with a conformal structure at the boundary.
 \subsection{Dirichlet problem for the boundary metric}
On the first place, the EGB action in Eq.~\eqref{Ibulk} needs to be supplemented by a boundary term to determine a well-posed Dirichlet problem at constant radius. It can be considered as a generalization of the Gibbons-Hawking-York term for Lovelock theory. In particular, for the Gauss-Bonnet scalar, the Myers term~\cite{Myers1987} can be cast as
\begin{equation}
I_{\rm Myers} = -\kappa \int\limits_{\partial \mathcal{M} }\diff{}^dx \sqrt{|h|}\;\beta_{\rm Myers}
\end{equation}
where $h=\det h_{ij}$ is the determinant of the induced metric in Gauss-normal coordinates and $\beta_{\rm Myers}$ is defined as
\begin{equation}
\beta_{\rm Myers}=2\delta^{i_1 i_2 i_3}_{j_1 j_2 j_3}K^{j_1}_{i_1}\left(\mathcal{R}^{j_2 j_3}_{i_2 i_3}-\frac{2}{3}K^{j_2}_{i_2}K^{j_3}_{i_3}\right).
\end{equation}
Here, $K_{ij}=-\tfrac{1}{2N}\partial_z h_{ij}$ is the extrinsic curvature in the Gauss-normal foliation (see Appendix~\ref{GC}). Furthermore, $\mathcal{R}^{ij}_{kl}(h)$ is the intrinsic curvature, which can be related to the bulk Riemann through the Gauss-Codazzi relation.

In the EGB case, the corresponding renormalized action the form is given by
\begin{align}
    I_{\rm EGB}^{\rm ren} &= \kappa\int\limits_{\mathcal{M} } \diff{}^{d+1}x \sqrt{|g|}\Big[R - 2\Lambda + \frac{\alpha}{4}\delta^{\mu_1...\mu_4}_{\nu_1...\nu_4}R^{\nu_1\nu_2}_{\mu_1\mu_2}R^{\nu_3\nu_4}_{\mu_3\mu_4} \Big] \notag \\
    & -2 \kappa \int\limits_{\partial \mathcal{M} } \diff{}^dx \sqrt{|h|}\left[K + 2\alpha \delta^{i_1i_2i_3}_{j_1j_2j_3}K^{j_1}_{i_1}\left(2 \mathcal{S}^{j_2}_{i_2}\delta^{j_3}_{i_3}-\frac{1}{3}K^{j_2}_{i_2}K^{j_3}_{i_3} \right) \right] + 2\kappa I_{\rm ct} \,,
    \label{regEGB}
\end{align}
where $I_{\rm ct}$ indicates a series of covariant boundary terms that cancel the divergences asymptotically. These can be determined in a systematic way, as  described for the first time in Ref.~\cite{deHaro:2000vlm}.
\subsection{Counterterms}
In practice, one first isolates the divergences at a finite cutoff, written in terms of the coefficients of the FG expansion. Then, everything is re-expressed as tensors of $g_{(0)ij}$. As a final step, one inverts the FG expansion in order to write the counterterms in terms of covariant tensors of the boundary metric $h_{ij}$. Although the method is, in principle, straightforward, the technical difficulty quickly increases with the bulk dimension and with the inclusion of higher-curvature terms, as is the case in Lovelock gravity.

The main steps of the derivation of the counterterms for EGB gravity are sketched below. Since the expansion to be considered is up to quadratic terms in the curvature, it is enough to present the inverted form of the curvature and determinant of the metric. In this case, all the quadratic terms can trivially be expressed as functions of $h_{ij}$. Indeed, taking into account Eqs.~\eqref{g2schouten} and~\eqref{g4},  one obtains 
\begin{align}
\sqrt{|g_{\left(0\right)}|} & = \frac{\sqrt{|h|} \ell_{\rm eff}^d}{ z^d}\left\{1+ \frac{z^2}{2} \mathcal{S}_{\left(0\right)} + \frac{z^4}{8} \left[ \mathcal{S}_{\left(0\right)ij}\mathcal{S}_{\left(0\right)}^{ij} + \mathcal{S}_{\left(0\right)}^2 + \frac{\alpha }{\ell^2_{\rm eff}\Delta^{(1)} \left(d-1 \right)} |\mathcal{W}_{(0)}|^2\right]+\mathcal{O} \left(z^6\right)\right\} \,, \label{detg0}\\
\mathcal{S}_{\left(0\right)}&= \frac{\ell_{\rm eff}^2}{ z^2} \left\{ \mathcal{S} \left(h\right) - \frac{1}{2\left(d-1\right)} \left[\left(d-2\right) \mathcal{S}_{ij} \left(h\right) \mathcal{S}^{ij} \left(h\right)+\mathcal{S}^2 \left(h\right)\right] +\mathcal{O} \left(\mathcal{R}^3\right)\right\} \,.\label{s0}
\end{align}
In order to track down the terms that render the action~\eqref{regEGB} infinite, the radial cutoff at $z=\epsilon$, where $\epsilon>0$ is introduced. In particular, the leading-order divergence coming from the sum of the bulk and the Myers terms, reads
\begin{equation}
I_{\rm div}^{\rm (0)} = - \frac{ \sqrt{|g_{\left(0\right)}|} \left(d-1\right)\ell_{\rm eff}^{d-1}}{ \epsilon^d} \left[1-\frac{2 \alpha \left(d-2\right) \left(d-3\right)}{\ell_{\rm eff}^2}\right] +\mathcal{O} \left(\epsilon^{2-d}\right) \,.
\end{equation}
Inverting the series, one determines the corresponding counterterm
\begin{equation}
I_{\rm ct}^{(0)}= \frac{ \sqrt{|h|} \left(d-1\right)}{\ell_{\rm eff}} \left[1-\frac{2 \alpha \left(d-2\right) \left(d-3\right)}{\ell_{\rm eff}^2}\right] \,.
\end{equation}
Taking this term into account, the next-to-leading order divergence of the action~\eqref{regEGB}, is linear in the curvature
\begin{equation}
I_{\rm div}^{\rm (1)} = - \frac{ \sqrt{|g_{\left(0\right)}|} \left(d-1\right)\ell_{\rm eff}^{d-1}}{ \left(d-2\right)\epsilon^{d-2}} \left[1 + \frac{2 \alpha \left(d-2\right) \left(d-3\right)}{\ell_{\rm eff}^2}\right] \mathcal{S}_{\left(0\right)} +\mathcal{O} \left(\epsilon^{4-d}\right) \,,
\end{equation}
which can be compensated by the counterterm
\begin{equation}
I_{\rm ct}^{\rm (1)}= \frac{ \sqrt{|h|} \left(d-1\right) \ell_{\rm eff}}{\left(d-2\right)} \left[1 + \frac{2 \alpha \left(d-2\right) \left(d-3\right)}{\ell_{\rm eff}^2}\right] \mathcal{S} \left(h\right) \,.
\end{equation}
In turn, the divergence that is quadratic in the curvature adopts the form
\begin{equation}
I_{\rm div}^{\rm (2)} = \frac{ \sqrt{|g_{\left(0\right)}|} \ell_{\rm eff}^{d-3}}{ 2\left(d-4\right)\epsilon^{d-4}} \left\{\ell_{\rm eff}^2 \left[1 - \frac{6 \alpha \left(d-2\right) \left(d-3\right)}{\ell_{\rm eff}^2} \right] \delta_{j_1 j_2}^{i_1 i_2} \mathcal{S}_{\left(0\right) i_1}^{j_1} \mathcal{S}_{\left(0\right) i_2}^{j_2} - \alpha |\mathcal{W}_{(0)}|^2 \right\} + \mathcal{O} \left(\epsilon^{6-d}\right) \,.
\end{equation}
This contribution can be  canceled by the addition of the counterterm
\begin{equation}
I_{\rm ct}^{\rm (2)}= -\frac{ \sqrt{|h|} \ell_{\rm eff}}{2 \left(d-4\right)} \left\{\ell_{\rm eff}^2 \left[1 - \frac{6 \alpha \left(d-2\right) \left(d-3\right)}{\ell_{\rm eff}^2} \right] \delta_{j_1 j_2}^{i_1 i_2} \mathcal{S}_{ i_1}^{j_1} \left(h\right)\mathcal{S}_{i_2}^{j_2} \left(h\right) - \alpha |\mathcal{W} \left(h\right)|^2 \right\}\,.
\end{equation}
Proper identification of divergent parts in the gravity action, and the inversion of the series in order to express them as covariant functions of the boundary metric, are at the core of this renormalization method. Performing this procedure for the EGB action~\eqref{regEGB}, one finds the counterterms that cancel divergences up to fourth derivative order and which can be written as
\begin{align}\label{EGBcounter}
 I_{\rm ct} &= 2 \kappa\int\limits_{\partial \mathcal{M} } \diff{}^dx \sqrt{|h|}\Bigg[c_0 + c_1\mathcal{S}  -c_2\delta^{i_1i_2}_{j_1j_2}\mathcal{S}^{j_1}_{i_1}\mathcal{S}^{j_2}_{i_2} + c_3|\mathcal{W}|^2\Bigg] \,,
\end{align}
where $\mathcal{S}=h^{ij}\mathcal{S}_{ij}$ is the trace of the intrinsic Schouten tensor and
\begin{align}\label{EGBcountercoeff}
c_0 &= \frac{d-1}{\ell_{\rm eff}}\left(1-\frac{2\alpha(d-2)(d-3)}{3\ell_{\rm eff}^2}\right)\,, \notag \\
c_1 &= \ell_{\rm eff}\frac{d-1}{d-2}\left(1+\frac{2\alpha(d-2)(d-3)}{\ell_{\rm eff}^2}\right)\,, \notag \\
c_2 &= \frac{\ell_{\rm eff}^3}{2(d-4)}\left(1-\frac{6\alpha(d-2)(d-3)}{\ell_{\rm eff}^2}\right)\,, \notag \\
c_3 &= \frac{\alpha \ell_{\rm eff}}{2(d-4)} \,.
\end{align}
Besides the modification of the coefficients that appear in the standard counterterms for Einstein-AdS, there is an extra piece corresponding to the last term which is exclusive to EGB-AdS gravity. Its presence eliminates the divergent contribution coming from non-conformally flat boundaries in the presence of the Gauss-Bonnet term.
\subsection{Quasilocal stress tensor}
An arbitrary variation of the action~\eqref{regEGB} with counterterms~\eqref{EGBcounter} gives  
\begin{align}
    \delta I_{\rm EGB}^{\rm ren} = \frac{1}{2}  
\int\limits_{\partial\mathcal{M}}\diff{}^dx\sqrt{|h|}\,\, T^{ij}\delta h_{ij}\,,
\end{align}
when the field equations hold.
The quasilocal stress tensor, in this basis of conformal quantities, is found to be
\begin{align}
 T^i_j[h] &= \frac{1}{8\pi G}\Bigg[\left(K^i_j - K\delta^i_j\right)  - 2\alpha \delta^{i_1...i_3 i}_{j_1...j_3 j}K^{j_1}_{i_1}\left(2 \mathcal{S}^{j_2}_{i_2}\delta^{j_3}_{i_3}-\frac{1}{3}K^{j_2}_{i_2}K^{j_3}_{i_3}\right) + c_0 \delta^i_j  \notag \\
 & -\frac{\left(d-2\right) c_1}{\left(d-1\right)} \left(\mathcal{S}^i_j - \mathcal{S} \delta^i_j  \right) + \frac{(d-4)\,c_2}{(d-2)}\Bigg(  \delta^i_j(\mathcal{S}^a_b \mathcal{S}^b_a - \mathcal{S}^2) - 2\mathcal{S}^i_a \mathcal{S}^a_j + 2\mathcal{S} \mathcal{S}^i_j -\frac{2}{d-4}\mathcal{B}^i_j \Bigg)  \notag \\
 & +c_3 \Big(\frac{1}{4}\delta^{i_1...i_4 i}_{j_1...j_4 j}\mathcal{W}^{j_1 j_2}_{i_1 i_2}\mathcal{W}^{j_3 j_4}_{i_3 i_4}-8(d-3)\mathcal{B}^i_j + 8(d-4)\mathcal{S}^a_b \mathcal{W}^{ib}_{ja}\Big) \Bigg]\, \label{Tmunu}
\end{align}
which is renormalized up to fourth derivative order, i.e., up to seven spacetime dimensions. A suitable rescaling in the radial holographic coordinate determines the vacuum expectation value of the holographic stress tensor as a given limit in Eq.~\eqref{Tmunu}, that is,
\begin{equation}\label{holoTmunu}
    \langle T_{ij} \rangle = \frac{2}{\sqrt{|g_{(0)}|}} \frac{\delta I_{\rm ren}}{\delta g_{(0)}^{ij}} = \lim_{z \to 0} \left( \frac{\ell_{\rm eff}^{d-2}}{z^{d-2}} T_{ij}[h] \right) \,.
\end{equation}
This gives the answer for the holographic energy-momentum tensor  for CFT$_5$, which takes the generic form
\begin{equation}
 \langle T_{ij} \rangle = -\frac{5}{16 \pi G_N \ell_{\rm eff}}   \Delta^{\left(1\right)}  g_{\left(5\right)ij} \,.
 \label{T5DHR}
\end{equation}

\section{Conformal Invariants and Renormalization}\label{Section 4}

A partial renormalization of EGB-AdS gravity can be achieved in even spacetime dimensions by the addition of a topological term of the Euler class. Indeed, this fully covariant contribution may be added on top of the six-dimensional action principle Eq.~\eqref{Ibulk}, without modifying its equations of motion
\begin{align}\label{action+ct}
    I^{\rm bulk} [g_{\mu\nu}] &= \kappa\int\limits_{\mathcal{M}_6}\diff{^6x}\sqrt{|g|}\left(R-2\Lambda + \alpha\mathcal{G} + \eta\mathcal{E}_6\right)\,.
\end{align}
The term $\mathcal{E}_6$ is defined as the six-dimensional Euler density
\begin{equation}\label{E6}
 \mathcal{E}_6 = \frac{1}{8}\delta^{\mu_1\ldots\mu_6}_{\nu_1\ldots\nu_6}R^{\nu_1\nu_2}_{\mu_1\mu_2}R^{\nu_3\nu_4}_{\mu_3\mu_4}R^{\nu_5\nu_6}_{\mu_5\mu_6} \,,
\end{equation}
whose coupling $\eta$ is fixed by requiring 
that the total action is zero for global AdS space~\cite{Kofinas:2006hr}
\begin{align}\label{eta1}
   \eta = -\frac{\ell_{\rm eff}^4}{72}\left(1-\frac{24\alpha}{\ell_{\rm eff}^2} \right)\,. 
\end{align}
N.B. the ambiguity on this prescription, given by any boundary term that depends on the bulk Weyl tensor, as the action would still be zero for the AdS vacuum.

Remarkably, for AAdS black-hole solutions of the theory, the action is finite and reproduces standard black hole thermodynamics. However, as anticipated above, the renormalization is partial because it is only reached in the case of conformally flat boundaries. Explicit computations below in the FG frame are intended to show this.

In order to do so, the bulk action of Eq.~\eqref{action+ct} can be conveniently written as a polynomial in $F$~\cite{Anastasiou:2021jcv}
\begin{align}
I^{\rm bulk}
    &=\frac{\kappa\ell_{\rm eff}^2}{48}\int\limits_{\mathcal{M}_6} \diff{^6}x\sqrt{|g|}\Bigg[\left(1-\frac{12\alpha}{\ell_{\rm eff}^2} \right)\delta^{\mu_1 ...\mu_4}_{\nu_1 ...\nu_4}F^{\nu_1 \nu_2}_{\mu_1 \mu_2} F^{\nu_3 \nu_4}_{\mu_3 \mu_4} \notag \\ &-\frac{\ell_{\rm eff}^2}{12}\left(1-\frac{24\alpha}{\ell_{\rm eff}^2} \right)\delta^{\mu_1 ...\mu_6}_{\nu_1 ...\nu_6}  F^{\nu_1 \nu_2}_{\mu_1 \mu_2} F^{\nu_3 \nu_4}_{\mu_3 \mu_4} F^{\nu_5 \nu_6}_{\mu_5 \mu_6}\Bigg]\,.\label{Ibulkren}
\end{align}
Notice that this action has a structure similar to that in Einstein-AdS$_6$ gravity, renormalized by adding a topological term \cite{Miskovic:2014zja}. 

In particular, the Pfaffian of the $F^{\mu\nu}_{\lambda\rho}$ can be radially decomposed as
\begin{align}\label{PfF}
\text{Pf}(F):=\delta^{\mu_1 ...\mu_6}_{\nu_1 ...\nu_6} F^{\nu_1 \nu_2}_{\mu_1 \mu_2} F^{\nu_3 \nu_4}_{\mu_3 \mu_4} F^{\nu_5 \nu_6}_{\mu_5 \mu_6}  = 12 \delta^{i_1 ...i_5}_{j_1 ...j_5} \left(F^{z j_{1}}_{z i_{1}} F^{j_{2} j_{3}}_{i_{2} i_{3}} F^{j_{4} j_{5}}_{i_{4} i_{5}} + 2 F^{z j_{1}}_{i_{1} i_{2}} F^{j_{2} j_{3}}_{z i_{3}} F^{j_{4} j_{5}}_{i_{4} i_{5}}\right) \,.
\end{align}
Taking into account the asymptotic expansion of Eq.~\eqref{Ftensorcompon} for a non-degenerate vacuum, the Pfaffian falls off as $\mathcal{O}\left(z^8\right)$, namely, faster than the normalizable modes in 6D. Therefore, this term does not contribute to the divergent structure of the action. Thus, the divergences should come only from the $F$-squared term, that is,
\begin{equation}
\delta^{\mu_1 ...\mu_4}_{\nu_1 ...\nu_4} F^{\nu_1 \nu_2}_{\mu_1 \mu_2} F^{\nu_3 \nu_4}_{\mu_3 \mu_4} = \delta^{i_1 ...i_4}_{j_1 ...j_4} F^{j_1 j_2}_{i_1 i_2} F^{j_3 j_4}_{i_3 i_4} + 8 \delta^{i_1 i_2 i_3}_{j_1 j_2 j_3}\left( F^{z j_1}_{z i_1} F^{j_2 j_3}_{i_2 i_3} + F^{j_1 z}_{i_1 i_2}  F^{j_2 j_3}_{i_3 z} \right) \,.
\end{equation}
For a generic AAdS space in EGB-AdS gravity, one may notice that the second and third terms of the above expression fall off at the normalizable order, producing no divergences whatsoever. Indeed, infinities are produced by the first term, which behaves asymptotically~as 

\begin{equation}
\delta^{i_1 ...i_4}_{j_1 ...j_4} F^{j_1 j_2}_{i_1 i_2} F^{j_3 j_4}_{i_3 i_4} = 4 \frac{z^4}{\ell_{\rm eff}^4} \mathcal{W}^{ms}_{\left(0\right) ab} \mathcal{W}^{ab}_{\left(0\right) ms} +\mathcal{O} \left(z^6\right) \,.
\label{Fsqasymptoics}
\end{equation}
Therefore, for Weyl-flat boundaries, one concludes that the $F$-squared term is finite. In that particular case, the addition of a topological term is enough to renormalize the action.\footnote{Notice that when the two vacua coalesce, the resulting combination of Eq.~\eqref{Fsqasymptoics} cannot be identified with any known geometrical object as $g_{\left(2\right) ij}$ remains undetermined.}

In order to define a renormalized action for generic AAdS spaces, one requires additional boundary terms. In order to avoid any ad-hoc prescription,  these counterterms are found by demanding bulk covariance and dependence on the Weyl tensor (as they should vanish for global AdS). Furthermore, they appear as a part of a conformal invariant in six dimensions. To start, a proper basis in that case is given by the invariants $\left\{I_1,I_2,I_3\right\}$ introduced in Ref.~\cite{Bastianelli:2000rs}, given by
\begin{align}
I_1&\equiv W_{\alpha\beta\gamma\delta}W^{\alpha\lambda\eta\beta}W_{\lambda}{}^{\gamma\delta}{}_{\eta}\, , \label{I1}\\
I_2&\equiv W_{\alpha\beta\gamma\delta}W^{\gamma\delta\lambda\eta}W_{\lambda\eta}{}^{\mu\nu}\, ,\label{I2}\\
I_3&\equiv W_{\alpha\gamma\delta\lambda}\left(\delta_\beta^\alpha\Box+4R^\alpha_\beta-\frac{6}{5}\delta^\alpha_\beta R\right)W^{\beta\gamma\delta\lambda}+\text{b.t.} ,\label{I3}
\end{align}
where b.t. denotes a total derivative. As one is interested in making contact with Lovelock gravity, defined in terms of totally anti-symmetric products of curvature tensors (Euler densities in all even dimensions lower), there is a symmetry enhancement for the combination
\begin{equation}
\mathcal{I}_{1}=32\left(2I_1+I_2\right)=\delta^{\mu_1 ...\mu_6}_{\nu_1 ...\nu_6}W^{\nu_1 \nu_2}_{\mu_1 \mu_2} W^{\nu_3 \nu_4}_{\mu_3 \mu_4} W^{\nu_5 \nu_6}_{\mu_5 \mu_6}:=\text{Pf}(W).
\end{equation}
Indeed, this term has a Lorentz symmetry in the tangent space, in addition to diffeomorphic invariance, when written in differential form language.

In a similar fashion, one may consider another combination of interest
\begin{align}\label{calI_2}
\mathcal{I}_{2}&=\frac{8}{3}I_1-\frac{2}{3}\left(I_2+I_3\right)\,,\nonumber \\
&=\delta^{\mu_1\cdots\mu_5}_{\nu_1\cdots\nu_5}W^{\nu_1\nu_2}_{\mu_1\mu_2}W^{\nu_3\nu_4}_{\mu_3\mu_4}S^{\nu_5}_{\mu_5} + 16|C|^2+\nabla_\mu \left( 16W^{\mu\nu\alpha\beta}C_{\nu\alpha\beta} - 2W^{\alpha\beta}_{\lambda\sigma}\nabla^\mu W^{\lambda\sigma}_{\alpha\beta}\right),
\end{align}
which, in a first-order formulation, incorporates the Hodge dual on top of Lorentz symmetry. For the later analysis, this invariant is conveniently split as $\mathcal{I}_{2}=\mathcal{I}^{\rm bulk}_{2}+\mathcal{I}^{\rm bdry}_{2}$.

The reduced basis  $\left\{\mathcal{I}_1,\mathcal{I}_2\right\}$ of conformal invariants has a couple of notable features. On one hand, both invariants are expected to be finite on their own when evaluated in the FG frame. In particular, the finiteness of $\text{Pf}(W)$ for asymptotically hyperbolic spaces has been discussed in the mathematical literature~\cite{Albin:2005qka}. In the second invariant, the boundary term is responsible for the cancellation of the infrared divergences coming from the bulk. On the other hand, a linear combination  $a_1\mathcal{I}_1+a_2\mathcal{I}_2$ for $a_1=\tfrac{1}{144}$ and $a_2=1$ corresponds to the conformal completion of Einstein-AdS action in six dimensions~\cite{Anastasiou:2020mik,Anastasiou:2023oro}. This is a Conformal Gravity theory with an Einstein sector, where the counterterms that appear are dictated by the conformal invariance of the action.

What is shown below is that a different choice of  $\left\{a_1,a_2\right\}$ is equivalent to the action of EGB-AdS on the holographic order. Therefore, its form gives insight on the type of fully covariant boundary terms required by the renormalization.

As a matter of fact, the difference between the Weyl tensor and the AdS curvature is of order $\mathcal{O}\left(z^4\right)$ in the asymptotic expansion,
\begin{equation}
W^{\mu\nu}_{\rho\lambda}= F^{\mu\nu}_{\rho\lambda} + \mathcal{O}(z^4)\,.
\end{equation}
Then, the expression in Eq.~\eqref{PfF} is given by
\begin{equation}
   \text{Pf}(F)=\text{Pf}(W)+\mathcal{O}\left(z^8\right) \,.
\end{equation}
For the evaluation of the bulk part of the other invariant, $\mathcal{I}_2$, it is useful to work out the expansion of the Schouten tensor
\begin{equation}
    S_{\lambda}^{\nu}= -\frac{1}{2\ell^2_{\rm eff}}\delta^\nu_\lambda + \mathcal{O}(z^4)\,, 
\end{equation}
and, from Eq.~\eqref{lqc}, the Cotton-squared contribution 
\begin{equation}
C^{\mu}_{\nu \rho}C^{\nu \rho}_{\mu}=\left\vert C \right\vert^2\sim\mathcal{O}\left(z^8\right).
\end{equation}
Therefore, the explicit evaluation in the FG asymptotic form of the metric leads to
\begin{equation}
\delta^{\mu_1 ...\mu_4}_{\nu_1 ...\nu_4}F^{\nu_1 \nu_2}_{\mu_1 \mu_2} F^{\nu_3 \nu_4}_{\mu_3 \mu_4}=-\ell^2_{\rm{eff}} \mathcal{I}_{2}^{\rm{bulk}}+\mathcal{O} \left( z^6 \right),
\end{equation}
such that it is suggestive to consider that the divergences of this term are canceled by the boundary part of the conformal invariant $-\ell^{2}_{\rm{eff}}\mathcal{I}_{2}^{\rm{bdry}}$. 
Finally, since
\begin{equation}
   \nabla_\mu \left(W^{\mu\nu\alpha\beta}C_{\nu\alpha\beta}\right) \sim \mathcal{O}\left(z^6\right) 
\end{equation}
is already of sub-normalizable order, one can further simplify $\mathcal{I}_{2}^{\rm{bdry}}$ such that the divergent piece \eqref{Fsqasymptoics} is expected to disappear thanks to the boundary term

\begin{equation} \label{DJmu}
I^{\rm bdry} =\int\limits_{\mathcal{M}_6}d^6x\sqrt{|g|}\nabla_\mu J^\mu\,,
\end{equation}
where the corresponding vector is given by
\begin{equation}\label{Jmu}
J^\mu=\frac{\kappa\ell^4_{\rm eff}}{24}\left(1-\frac{12\alpha}{\ell^2_{\rm eff}} \right)W^{\alpha\beta}_{\lambda\sigma}\nabla^\mu W^{\lambda\sigma}_{\alpha\beta}  = \frac{\kappa\ell_{\rm eff}^4\Delta^{(1)}}{24} W^{\alpha\beta}_{\lambda\sigma}\nabla^\mu W^{\lambda\sigma}_{\alpha\beta}  .
\end{equation}
The renormalization of the six-dimensional EGB-AdS action is explicitly verified in the next section, when the above formula is used as a fully covariant counterterm.

\section{Fully covariant counterterms in EGB AdS gravity in 6D\label{sec:CR}}

In even-dimensional Einstein-AdS gravity, bulk counterterms are given by the addition of the Euler density with a fixed coupling constant, which renders the total action finite for asymptotically conformally flat manifolds~\cite{Olea:2005gb,Miskovic:2009bm,Anastasiou:2018rla, Anastasiou:2018mfk, Anastasiou:2020zwc}. Nevertheless, additional divergences will appear if the asymptotic boundary is not Weyl-flat. A different renormalization scheme was later introduced, which complements the previous prescription and accounts for generic AAdS spacetimes~\cite{Anastasiou:2020mik, Anastasiou:2022ljq, Anastasiou:2022wjq, Anastasiou:2023oro}.

Interestingly, the use of topological densities of the Euler class as counterterms is not restricted only to Einstein-AdS gravity. In particular, it was shown that the same prescription turns EGB-AdS~\cite{Kofinas:2006hr} and, in general, Lovelock-AdS~\cite{0708.0782,Kofinas:2008ub} gravity free from IR divergences for asymptotically conformally flat spaces.  Nevertheless, its extension for generic AAdS spacetimes is still lacking. This section is intended to shed some light on this problem.

Consider the action of Eq.~\eqref{action+ct} with the extra boundary term, 
\begin{align}\label{action+J}
    I_{\rm EGB}^{\rm ren} [g_{\mu\nu}] &= I^{\rm bulk} +  \int\limits_{{\mathcal{M}_6}}\diff{^6x}\sqrt{|g|} \nabla_{\mu} J^{\mu} \,,
\end{align}
where $J^{\mu}$ is given by Eq.~\eqref{Jmu}, and which can be equivalently written~as 
\begin{align}\label{Jmuz}
J^\mu &=\frac{\kappa\ell^4_{\rm eff}\Delta^{(1)}}{24}F^{\alpha\beta}_{\lambda\sigma}\nabla^\mu F^{\lambda\sigma}_{\alpha\beta}+\mathcal{O}\left(z^9\right)\,.
\end{align}
This is the boundary term required to cancel the remaining divergence coming from Eq.~\eqref{Fsqasymptoics}.
Indeed, the asymptotic value of this term differs from 
\begin{align}\notag
\int\limits_{\partial \mathcal{M}_6} \diff{^5x}\sqrt{|h|} \, n_{\mu}F^{\alpha\beta}_{\lambda\sigma}\nabla^\mu F^{\lambda\sigma}_{\alpha\beta}&=\frac{1}{8}\int\limits_{\partial \mathcal{M}_6} \diff{^5x}\sqrt{|h|} \, n_{\mu} \nabla^{\mu} \left(\delta^{\mu_1 ...\mu_4}_{\nu_1 ...\nu_4} F^{\nu_1 \nu_2}_{\mu_1 \mu_2} F^{\nu_3 \nu_4}_{\mu_3 \mu_4}\right) + \mathcal{O}\left(z\right) \\
&= - \frac{1}{2z}   \int\limits_{\partial \mathcal{M}_6}\diff{^5x}\sqrt{|g_{\left(0\right)}|}\, \mathcal{W}^{ij}_{\left(0\right) ms} \mathcal{W}^{ms}_{\left(0\right) ij} + \mathcal{O} \left(z\right) \,,
\end{align}
by contributions of sub-holographic order, as it involves traces of $F$.
Therefore, the renormalized Einstein-Gauss-Bonnet can be expressed solely in terms of the AdS curvature as 
\begin{align}\notag
    I_{\rm{EGB}}^{\rm{ren}} [g_{\mu\nu}]&=\kappa\int_{\mathcal{M}_6} \diff{^6}x\sqrt{|g|}\Bigg(\frac{\eta}{8}\delta^{\mu_1 ...\mu_6}_{\nu_1 ...\nu_6} F^{\nu_1 \nu_2}_{\mu_1 \mu_2} F^{\nu_3 \nu_4}_{\mu_3 \mu_4} F^{\nu_5 \nu_6}_{\mu_5 \mu_6}+ \frac{\ell_{\rm eff}^2 \Delta^{(1)} }{48}\delta^{\mu_1 ...\mu_4}_{\nu_1 ...\nu_4} F^{\nu_1 \nu_2}_{\mu_1 \mu_2} F^{\nu_3 \nu_4}_{\mu_3 \mu_4} \Bigg) \\
    & + \int\limits_{\partial\mathcal{M}_6}\diff{^5}x\sqrt{|h|}\,n_\mu J^\mu\,.
    \label{Iren}
\end{align}

\subsection{Variation of a covariant renormalized action}

The different form of the renormalized action~\eqref{Iren} for EGB-AdS gravity in six spacetime dimensions makes possible an alternative derivation of the holographic stress tensor.
In order to simplify the computation of this tensor, it is convenient to decompose the variation into two parts, such that
\begin{equation}
\delta I_{\rm{EGB}}^{\rm{ren}} = \delta I^{\rm bulk} + \delta I^{\rm bdry} \,.
\end{equation}
First, an arbitrary on-shell variation of the bulk term produces
\begin{equation}
    \delta I^{\rm bulk} =  -\kappa \int\limits_{\partial \mathcal{M}_6}d^5 x\sqrt{|h|} \delta^{\mu_1 ...\mu_6}_{\nu_1 ....\nu_6}  F^{\nu_1 \nu_2}_{\mu_1 \mu_2}\delta \Gamma^{\nu_3}_{\lambda \mu_3} g^{\nu_4 \lambda} n_{\mu_4} \left(\frac{3\eta}{4} F^{\nu_5 \nu_6}_{\mu_5 \mu_6}  +\frac{\ell_{\rm eff}^2\Delta^{(1)}}{48} \delta^{\nu_5 \nu_6}_{\mu_5\mu_6} \right) \,,
    \label{varEGBbulkcov}
\end{equation}
which can be cast in the form
\begin{align}
    \delta I^{\rm bulk} &= -\kappa \int\limits_{\partial \mathcal{M}_6}d^5 x\sqrt{|h|}\left\{\frac{3\eta}{4}\delta^{i_1 ...i_5}_{j_1 ...j_5} F^{j_1 j_2}_{i_1 i_2} \Big[F^{j_3 j_4}_{i_3 i_4}(2\delta K^{j_5}_{i_5}+K^m_{j_5}(h^{-1}\delta h)^{j_5}_m)   \right. \notag \\
    & \left. + 8\mathcal{D}^{j_3}\mathcal{D}_{i_3}K^{j_4}_{i_4}(h^{-1}\delta h)^{j_5}_{i_5}) \Big] +\frac{\ell_{\rm eff}^2\Delta^{(1)}}{12}\delta^{i_1 i_2 i_3}_{j_1 j_2 j_3}\Big[F^{j_1 j_2}_{i_1 i_2}(2\delta K^{j_3}_{i_3}+K^m_{j_3}(h^{-1}\delta h)^{j_3}_m)\right. \notag \\
    &\left. -4\mathcal{D}^{j_1}\mathcal{D}_{i_1}K^{j_2}_{i_2}(h^{-1}\delta h)^{j_3}_{i_3})\Big]\right\}\,,
\end{align}
where $\mathcal{D}$ is the Levi-Civita covariant derivative compatible with $h_{ij}$. 
Unlike with the HR prescription given in Sec.~\ref{S4}, the resulting variation does not define a quasilocal stress energy tensor at a finite radial cutoff. Instead, it is expected to obtain a well-defined variational principle for the boundary metric $g_{(0) ij}$, where the holographic stress tensor can be read off. To do so, the FG expansion is plugged in the above variation, which gives
\begin{align}\label{VarBulk}
    \delta I^{\rm bulk} = - \frac{\kappa\ell_{\rm eff}^4\Delta^{(1)}}{24}  \int\limits_{\partial \mathcal{M}_6}d^5 x\sqrt{|g_{(0)}|}\Bigg[\frac{1}{z}\bigg(\frac{1}{4}\delta^{i_1 ...i_5}_{j_1 ...j_5}\mathcal{W}^{(0)j_1 j_2}_{i_1 i_2}\mathcal{W}^{(0)j_3 j_4}_{i_3 i_4}(g^{-1}_{(0)}\delta g_{(0)})^{j_5}_{i_5} \notag \\ - 8(\mathcal{B}^{(0)i}_j+\mathcal{D}^{(0)m}\mathcal{C}^{(0)i}_{jm})(g^{-1}_{(0)}\delta g_{(0)})^{j}_{i}\bigg)+\frac{60}{\ell^5_{\rm eff}}g^i_{(5)j}(g^{-1}_{(0)}\delta g_{(0)})^{j}_{i}+\mathcal{O}(z) \Bigg] \,.
\end{align}
A proper holographic description gives rise to finite contributions at the conformal boundary, in terms of variations of the holographic source. Notice that the finiteness of this variational problem for $g_{(0)ij}$ is spoiled for generic boundary geometry. In that respect, the addition of the Euler density in 6D with fixed coupling is insufficient in general.

On top of the latter, the total derivative of Eq.~\eqref{action+J} resolves the mismatch without contributing at finite order, as it can be seen below. Its generic variation reads
\begin{equation}
  \delta I^{\rm bdry} =  \int\limits_{\partial \mathcal{M}_6}d^5 x\sqrt{|h|}n^\mu \left[\frac{1}{2} J_\mu (g^{-1}\delta g) + \delta J_\mu - J_\nu (g^{-1}\delta g)^\nu_\mu \right] \,,
\end{equation}
which can be cast in the form
\begin{equation}
    \delta I^{\rm bdry}
    = \frac{\kappa\ell_{\rm eff}^3\Delta^{(1)}}{384} \int\limits_{\partial \mathcal{M}_6}d^5 x\sqrt{|h|} \delta^{\mu_1 ...\mu_4}_{\nu_1 ...\nu_4} z \left[\partial_z \left(F^{\nu_1 \nu_2}_{\mu_1 \mu_2}F^{\nu_3 \nu_4}_{\mu_3 \mu_4} \right)(h^{-1}\delta h)+2 \partial_z \delta \left(F^{\nu_1 \nu_2}_{\mu_1 \mu_2}F^{\nu_3 \nu_4}_{\mu_3 \mu_4}\right) \right] \,. \label{deltaIbdry}
\end{equation}

\subsection{Holographic stress tensor}

In the FG gauge, the expression in Eq.~\eqref{deltaIbdry} is expanded asymptotically as
\begin{align}\label{VarBdry}
    \delta I^{\rm bdry} = -\frac{\kappa\ell_{\rm eff}^4\Delta^{(1)}}{24}  \int\limits_{\partial \mathcal{M}_6}d^5 x\sqrt{g_{(0)}}\Bigg[\frac{1}{z}\bigg(\frac{1}{4}\delta^{i_1 ...i_5}_{j_1 ...j_5}\mathcal{W}^{(0)j_1 j_2}_{i_1 i_2}\mathcal{W}^{(0)j_3 j_4}_{i_3 i_4}(g^{-1}_{(0)}\delta g_{(0)})^{j_5}_{i_5} \notag \\ - 8(\mathcal{B}^{(0)i}_j+\mathcal{D}^{(0)m}\mathcal{C}^{(0)i}_{jm})(g^{-1}_{(0)}\delta g_{(0)})^{j}_{i}\bigg) + \mathcal{O}(z) \Bigg] \,.
\end{align}
Hence, the total derivative contribution in Eq.\eqref{DJmu} is the minimal subtraction of topological renormalization for EGB-AdS gravity. Indeed, it is straightforward to show
that the total variation of the action reads
\begin{align}\label{Thol}
     \delta I_{\rm EGB}^{\rm ren}  = \frac{1}{2}\int\limits_{\partial \mathcal{M}_6}\diff{^5x}\sqrt{|g_{(0)}|}\delta g_{(0)}^{ij}\langle T_{ij} \rangle\,, \;\;\;\;\; \mbox{where} \;\;\;\;\; \langle T_{ij} \rangle = -\frac{5\Delta^{(1)}}{16\pi G\,\ell_{\rm eff}} \, g_{(5)ij}
\end{align}
is the holographic stress tensor, which matches exactly that obtained via  Holographic Renormalization [cf. Eq.~\eqref{T5DHR}].

\section{Six-dimensional topological Boulware-Deser black holes}\label{S6}

In order to apply the different versions of the renormalized action of the theory, one may consider asymptotically AdS static black holes in Einstein-Gauss-Bonnet theory. To this end, one may focus on static metrics  
\begin{align}\label{metricansatz}
    \diff{s^2} = -f(r)\diff{t^2} + \frac{\diff{r^2}}{f(r)} + r^2\diff{\Sigma_{(4)}}^2\,, \;\;\;\;\; \mbox{where} \;\;\;\;\; \diff{\Sigma_{(4)}}^2 = \sigma_{mn}(\varphi)\diff{\varphi^m}\diff{\varphi^n}\,,
\end{align}
is an Einstein metric of a codimension-2 hypersurface parametrized by local coordinates~$\{\varphi^m\}$. 

\subsection{Boundary topology}
In this case, the topology of the transversal section is of a different sort, as black holes with $\mathbb{S}^4$, $\mathbb{H}^4$, $\mathbb{T}^4$,  $\mathbb{S}^2\times\mathbb{S}^2$, $\mathbb{H}^2\times\mathbb{H}^2$, $\mathbb{CP}^2$ and $\mathbb{CH}^2$ horizon can be found. While the first three are conformally flat, the others are not. Therefore, they provide a suitable setup for studying CFTs on more general backgrounds, where infrared divergences produced by non-asymptotically conformally flat boundaries of topological black holes can still be renormalized. For instance, the $\mathbb{CP}^2$ metric is a Weyl self-dual Euclidean Einstein manifold whose Fubini-Study metric can be parametrized as
\begin{align}\label{CPK}
 \diff{\Sigma_{(4)}^2} &= 6\bigg[\diff{\psi_2^2} + \sin^2\psi_2\cos^2\psi_2\left(\diff{\phi_2} + \sin^2\psi_1\diff{\phi_1} \right)^2 
 + \sin^2\psi_2\diff{\Omega^2} \bigg]\,,
 \end{align}
where $\diff{\Omega^2}$ is the metric of the round two-sphere. The $\mathbb{CH}^2$ space, on the other hand, admits a similar fibration as $\mathbb{CP}^2$. Indeed, its line element can be obtained from Eq.~\eqref{CPK} by replacing $\cos\psi_2\to\cosh\psi_2$ and $\sin\psi_2\to\sinh\psi_2$.

Replacing the line element~\eqref{metricansatz} in the Einstein-Gauss-Bonnet field equations~\eqref{eom}, one finds that they are solved by the metric function~\cite{boulware1985}
\begin{align}\label{fsol}
    f_\pm(r) = k + \frac{r^2}{12\alpha}\left[1 \pm \sqrt{\left(1+\frac{12\alpha\Lambda}{5} \right)-\frac{16(1+\sigma)\alpha^2}{r^4}+\frac{96\pi\alpha mG}{\Sigma_{(4)}r^5}}\,\right]\,,
\end{align}
where $m$ is an integration constant associated to the mass, $\Sigma_{(4)}$ is the volume of the transverse section, while $k$ and $\sigma$ are related to its topology. Their specific values are summarized in Table~\ref{tabla1}. If the topology of the horizon is $\mathbb{S}^4$, this solution describes the one first found by Boulware and Deser in Ref.~\cite{boulware1985}. Thus, the solution in Eq.~\eqref{fsol} will be referred to as Boulware-Deser topological black holes. 

\begin{table}[h]
\begin{centering}
\begin{tabular}{cccccccc}
\hline
 & $\;\;\;\mathbb{S}^4\;\;\;$ & $\;\;\;\mathbb{T}^4\;\;\;$ & $\;\;\;\mathbb{H}^4\;\;\;$ & $\;\;\;\mathbb{CP}^2\;\;\;$ & $\;\;\;\mathbb{CH}^2\;\;\;$ & $\;\mathbb{S}^2\times\mathbb{S}^2\;$ & $\;\mathbb{H}^2\times\mathbb{H}^2\;$ \\
 \hline
 $\;\;\;k\;\;\;$ & $1$ & $0$ & $-1$ & $1/3$ & $-1/3$ & $1/3$ & $-1/3$ \\
 \hline
 $\;\;\;\sigma\;\;\;$ & $-1$ & $-1$ & $-1$ & $0$ & $0$ & $1$ & $1$ \\
 \hline
\end{tabular}
\caption{Properties of transverse sections.}\label{tabla1}
\end{centering}
\end{table}

The metric~\eqref{metricansatz} with lapse function~\eqref{fsol} has a curvature singularity at the origin. The singularity, however, is hidden behind an event horizon located at the locus $r=r_+$, defined as the largest positive root of the polynomial $f(r_+)=0$. This definition allows one to express the $m$ in terms of $r_+$ as
\begin{align}
    m = \frac{\Sigma_{(4)}r_+}{4\pi G}\left[kr_+^2 + 6\alpha k^2 + \frac{2\alpha(\sigma+1)}{3} + \frac{r_+^4}{\ell_{\rm eff}^2} - \frac{6\alpha r_+^4}{\ell_{\rm eff}^4}  \right]\,.
\end{align}
\subsection{Asymptotic form}
The branch $f_-(r)$ in Eq.~\eqref{fsol} is continuously connected to topological black holes of general relativity in the limit $\alpha\to0$, since
\begin{align}\label{fsolm}
 f_-(r) = k - \frac{4\pi mG}{\Sigma_{(4)}r^3} +\frac{r^2}{\ell^2} +  \frac{1}{r^8}\left[\frac{2(\sigma+1)r^6}{3} + \frac{3\left(\Lambda\Sigma_{(4)}r^5+40\pi mG\right)^2}{50\Sigma_{(4)}^2}\right] \alpha +  \mathcal{O}(\alpha^2)\,,
\end{align}
where $\ell_{\rm eff}\to\ell$ as $\alpha\to0$, with $\ell_{\rm eff}$ being defined in terms of the cosmological constant via Eq.~\eqref{leff}. Thus, in this limit, it is clear that this solution becomes the Schwarzschild-Tangherlini black hole~\cite{Tangherlini:1963bw}. On the other hand, the solution $f_+(r)$ is nonperturbative in~$\alpha$. Henceforth, the analysis below focuses on the branch~\eqref{fsolm} for the sake of simplicity, denoted by $f_-(r)=f(r)$ from hereon. The nonperturbative case can be treated similarly. The asymptotic behavior of the metric function is
\begin{align}
 f(r) = \frac{r^2}{\ell_{\rm eff}^2} + k  + \left(\frac{\ell_{\rm eff}^2}{\ell_{\rm eff}^2-12\alpha}\right) \left(\frac{2\alpha(\sigma+1)}{3r^2} -\frac{4\pi mG}{\Sigma_{(4)}\,r^3} \right) + \mathcal{O}(r^{-6})\,,
\end{align}
when $r\to\infty$. Thus, it is clear that, in EGB AdS gravity, a non-conformally flat boundary ($\sigma\neq-1$)  contributes with an additional term in the asymptotic expansion, whose falloff is weaker than the Newtonian potential.

\section{Conserved charges and thermodynamics}\label{S7}

In order to compute the renormalized conserved charges for the topological Boulware-Deser-AdS black holes, the Noether-Wald formalism~\cite{Wald:1993nt, Iyer:1994ys} will be used. To this end, the Einstein-Gauss-Bonnet theory augmented by the fully covariant counterterms of Sec.~\ref{sec:CR} will be considered. 
  
\subsection{Noether-Wald charges}
The Noether-Wald formalism uses the fact that the current associated with diffeomorphism invariance generated by a Killing vector field $\xi=\xi^\mu\partial_\mu$ is conserved on shell. The Noether current can be written locally as an exact form, whose prepotential can be expressed by~\cite{Wald:1993nt,Iyer:1994ys}
\begin{align}
    q^{\mu\nu} = - \left(E^{\mu\nu}_{\lambda\rho}\nabla^\lambda\xi^\rho + 2\xi^\lambda\nabla^\rho E^{\mu\nu}_{\lambda\rho} \right)=-q^{\nu\mu}\,.
\end{align}
Here, $E^{\mu\nu}_{\lambda\rho}$ denotes the functional derivative of the bulk Lagrangian with respect to the Riemann tensor. For the action~\eqref{action+ct}, this quantity is given by
\begin{align}
    E^{\mu\nu}_{\lambda\rho} = \frac{\kappa}{2}\left(\delta^{\mu\nu}_{\lambda\rho} + \alpha\,\delta^{\mu\nu\mu_1\mu_2}_{\lambda\rho\nu_1\nu_2}R^{\nu_1\nu_2}_{\mu_1\mu_2} + \frac{3\eta}{4}\,\delta^{\mu\nu\mu_1\ldots\mu_4}_{\lambda\rho\nu_1\ldots\nu_4}R^{\nu_1\nu_2}_{\mu_1\mu_2}R^{\nu_3\nu_4}_{\mu_3\mu_4} \right)\,.
\end{align}
Then, the asymptotic Noether-Wald charges associated with a Killing vector field $\xi$ can be obtained from~\cite{Wald:1993nt,Iyer:1994ys}
\begin{align}\label{charges}
    Q[\xi] = \int\limits_{\Sigma_{\infty}}\left(q^{\mu\nu} - 2\xi^{[\mu}J^{\nu]} \right)\diff{\Sigma_{\mu\nu}} \,,
\end{align}
where $\diff{\Sigma_{\mu\nu}}$ is the volume element of the codimension-2 boundary, $\Sigma_\infty$ denotes spacelike infinity, and $J^\mu$ is defined in Eq.~\eqref{Jmu}.

The mass of the topological Boulware-Deser black hole can be obtained as a conserved charge associated with the asymptotically time-like Killing vector $\xi=\partial_t$. Direct evaluation of Eq.~\eqref{charges} gives
\begin{align}\label{mass}
   Q[\partial_t] := M = m\,.
\end{align}
\subsection{Quasilocal charges}
This result can be compared with that obtained via the renormalized quasilocal stress tensor in Eq.~\eqref{Tmunu}. To this end, the induced metric is written in an ADM form, that is,
\begin{equation}
    h_{ij} \, \diff{x^i} \diff{x^j} = - \tilde{N}^2 \diff{t^2} + \gamma_{mn} \left( \diff{y^m} + N^m \diff{t} \right) \left( \diff{y^n} + N^n \diff{t} \right)\,.
\end{equation}
Then, the quasilocal charges associated with the Killing vector $\xi=\xi^\mu\partial_\mu$ are defined as~\cite{Balasubramanian:1999re}
\begin{align} \label{Qquasilocal}
    \mathcal{Q}[\xi] = - \int\limits_{\Sigma_\infty}\diff{^{d-1}y}\sqrt{|\gamma|}\,T^i_j\,\xi^j\, u_i\,,
\end{align}
where $u^i$ is the timelike unit-normal to spacelike infinity and $\gamma$ is the determinant of its codimension-2 boundary metric. Considering the topological Boulware-Deser-AdS black hole discussed in Sec.~\ref{S6}, together with its asymptotically time-like Killing vector $\xi=\partial_t$, we find that its mass is $\mathcal{Q}[\partial_t]=m$, in full agreement with the Noether-Wald formalism. 

Additionally, the holographic stress tensor dual to this topological black hole can be computed directly from Eq.~\eqref{holoTmunu}, giving the following
\begin{align}\label{TmunuBD}
    \langle \, T^i_j \,\rangle = -\frac{m}{4\Sigma_{(4)}\ell_{\rm (eff)}^4}\begin{pmatrix} -4 & 0 \\ 0 & \mathbb{I}_{4x4} \end{pmatrix} \,.
\end{align}
In this form, it is clear that the holographic stress tensor is traceless, which is consistent with the fact that there is no conformal anomaly in odd (boundary) dimensions. One can check that this holographic stress tensor is thermal, as it can be written in a perfect-fluid form. Its energy density can be read off directly from Eq.~\eqref{TmunuBD}, being proportional to the mass of the black hole solution. Since the holographic stress tensor is traceless, it satisfies an equation of state for an ultra-relativistic fluid in five dimensions.

\subsection{Black hole Thermodynamics}
The entropy, on the other hand, is defined by the Noether-Wald charge evaluated at the bifurcating Killing horizon $\mathcal{H}$, that is,
\begin{align}\label{entropyNW}
 S = \frac{1}{T}\int\limits_{\mathcal{H}}q^{\mu\nu}\diff{\Sigma_{\mu\nu}}\,.
\end{align}
The Hawking temperature of the black hole is denoted by $T$ and it can be computed directly from $T=\tfrac{\kappa_s}{2\pi}$, where $\kappa_s$ is the surface gravity. Direct evaluation of Eq.~\eqref{entropyNW} alongside the surface gravity of the topological Boulware-Deser black hole, allows to obtain the entropy and temperature of the solution, which are given by
\begin{subequations}\label{thermostuff}
    \begin{align}
    \label{entropy}
    S &= \frac{r_+^4\Sigma_{(4)}}{4G}\left(1+\frac{24\alpha k}{r_+^2} \right) + S_0\,, \\
    \label{temperature}
    T &=  \frac{1}{12\pi r_+\left(r_+^2 + 12\alpha k \right)}\left( 9kr_+^2+2\alpha\left[9k^2+\sigma+1 \right] + \frac{15r_+^4}{\ell_{\rm eff}^2}  - \frac{90\alpha r_+^4}{\ell_{\rm eff}^4} \right)\,,
\end{align}
\end{subequations}
respectively. As is well-known, black hole entropy in Einstein-Gauss-Bonnet gravity does not follow the standard area law because it receives contributions from the higher-curvature terms~\cite{Jacobson:1993xs,Cai:2001dz}. Additionally, $S_0$ is a constant contribution to the entropy that does not affect the first law of thermodynamics. The latter has a topological origin, as it is related to the Euler characteristic of the black hole's horizon via
\begin{align}
    S_0 = \frac{24\pi^2\eta}{G}\,\chi(\Sigma_{(4)})\,,
\end{align}
where $\eta$ is defined in Eq.~\eqref{eta1}.

To first order in the saddle-point approximation, the partition function can be obtained directly from the renormalized Euclidean on-shell action via $\ln\mathcal{Z}\approx-I_E^{\rm ren}$. To compute the latter, we perform the analytic continuation $t\to-i\tau$ in Eq.~\eqref{metricansatz}. The absence of conical singularities at the horizon requires that the Euclidean time must be identified as $\tau\sim\tau+\beta$, whose period is $\beta=1/T$.

Then, from the fully covariant renormalized Einstein-Gauss-Bonnet-AdS action, one can compute the Helmholtz free energy for the Boulware-Deser black hole, which is given by $\mathcal{F}=\beta^{-1}I_E^{\rm ren}$. Thus, from the Euclidean on-shell action in Eq.~\eqref{Iren}, the free energy yields
\begin{align}\label{QSR}
    \mathcal{F} = M - TS\,,
\end{align}
in the canonical ensemble. Here $M$, $T$, and $S$, are the mass, temperature, and entropy of the black hole, respectively, which are exactly those obtained via the Noether-Wald formalism in Eqs.~\eqref{mass} and~\eqref{thermostuff}. Thus, from this expression, it is clear that these conserved charges satisfy the first law of thermodynamics.

\section{Discussion and conclusions}\label{S8}

In this work,  Holographic Renormalization is performed for Einstein-Gauss-Bonnet AdS gravity in arbitrary dimensions, up to order $z^5$ in the Poincaré coordinate, for a generic boundary metric $g_{(0)ij}$. In particular, this procedure deals with divergences in AAdS spaces induced by a non-conformally flat boundary metric. On purpose, the corresponding counterterms are expressed in terms of the Weyl, Schouten and Cotton tensors, quantities of interest in the context of Conformal Calculus. They adequately account for nontrivial conformal properties of the boundary. This result contrasts with existing literature~\cite{Brihaye:2008xu} where, assuming covariance at the boundary, the same counterterms series as for Einstein gravity was used. The coefficients of this series were then fixed by demanding finiteness of the action for ACF black holes, which is not the case here.  It is worthwhile to emphasize that the coefficients of the local counterterms in Eq.~\eqref{EGBcounter} are nonlinear in $\alpha$. When linearized in this parameter, the expressions match the ones obtained perturbatively by the Hamilton-Jacobi method in Ref.~\cite{Liu:2008zf}.

Topological black holes with boundaries with a nontrivial conformal structure also serve to explore modifications of renormalization methods in higher-curvature AAdS gravity. As these conformal properties are reflected in divergent terms in the action in six and higher dimensions, in the case of six-dimensional EGB-AdS, the action is renormalized by fully covariant counterterms (see Section \ref{sec:CR}). Said counterterms extend the topological renormalization result of Ref.~\cite{Kofinas:2006hr}, by considering an additional term, $\Box \lvert W \rvert^2$, which cancels a divergence that appears in generic AAdS manifolds.
 
When rewriting the bulk action in terms of a polynomial of the AdS curvature, it is easy to understand the role of $\Box \lvert W \rvert^2$. Basically, as shown in Eq.~\eqref{Iren}, the polynomial part of the renormalized action considers a quadratic and a cubic term in $F$. However, the cubic term is finite on its own. Therefore, the role of the extra counterterm is simply to cancel the divergence in the quadratic part of the action. This behavior mimics the situation for Einstein-AdS gravity when considering the Conformal Renormalization procedure, as shown in Ref.~\cite{Anastasiou:2023oro}. The reason behind this is the universality of next-to-leading order coefficient in the FG expansion~\cite{hep-th/9910267}. Indeed, $g_{(2)ij}$ is fixed in terms of the Schouten tensor of $g_{(0)ij}$, what is a generic property of any AAdS spacetime, that can be shown even at a kinematic level. Particular features of EGB AdS appear at order $z^4$, where $g_{(4)ij}$ portrays an additional $\mathcal{W}_{(0)}^2$ contribution proportional to the GB coupling. However, such deformation is not enough to modify the conformal invariant structure in Eq.~\eqref{calI_2} until holographic order. As a consequence, one can borrow the form of the total derivative, which renders the action finite.

As an application of this fully covariant renormalized action, Noether-Wald charges are derived, and explicitly computed for topological Boulware-Deser black holes. In doing so, the black hole mass, understood as internal energy, gets combined with the entropy and Helmholtz free energy to form the Quantum Statistical Relation~\eqref{QSR} (see \cite{Gibbons:2004ai}).
Regarding quasilocal methods,  the renormalized stress tensor in Eq.~\eqref{Tmunu} also gives the correct conserved quantities for the AAdS solutions mentioned above. The use of boundary counterterms also reproduces the correct black hole thermodynamics. The holographic energy-momentum tensor for this family of solutions is traceless, as there is no conformal anomaly in odd-dimensional boundaries. Indeed, its perturbations describe an  almost-conformal perfect fluid, where transport coefficients can be obtained in non-conformally flat backgrounds \cite{Kovtun:2004de}.

Renormalized gravity actions which are fully covariant in the bulk have a number of useful advantages, which have been explored to a certain extent. For instance, as shown in Ref.~\cite{Anastasiou:2021jcv}, the computation of the universal part of the holographic entanglement entropy for higher-curvature gravity theories~\cite{Dong:2013qoa}, such as Lovelock, simplifies considerably when the renormalized action is written in terms of spacetime-covariant objects. The implementation of $\Box \lvert W \rvert^2$ in codimension-2 would allow to obtain corrections to the Willmore energy or of the reduced Hawking mass in Einstein-AdS gravity~\cite{Anastasiou:2024rxe}. In addition to applications in holographic quantum information, in the context of the fluid/gravity correspondence, the covariant form of the renormalized EGB action may simplify the calculation of physical quantities in different coordinate frames~\cite{Rangamani:2009xk,Bhattacharyya:2007vjd}.

Both quasilocal and Noether-Wald charges are intended to amend another notion of conserved quantities in AAdS spacetimes, which is Conformal Mass~\cite{AshtekarMagnon1984,Ashtekar:1999jx}. In the case of EGB-AdS gravity, the Conformal Mass definition, is given by the surface integral~\cite{Pang:2011cs,Jatkar:2015ffa}
\begin{align} \label{AMDcharge}
    \mathcal{H_{\rm EGB}}[\xi] = -\frac{\ell_{\rm eff}\,\Delta^{(1)}}{8\pi G} \int\limits_{\Sigma_\infty}\diff{^{d-1}y}\sqrt{|\gamma|}\,\mathcal{E}^i_j\,\xi^j\, u_i\,,
\end{align}
where $\mathcal{E}^i_j$ is the electric part of the Weyl tensor, defined as
\begin{align} \label{electicW}
    \mathcal{E}^i_j = \frac{1}{d-2}\,W^{i\nu}_{j\mu}\,n^\mu\, n_\nu\,,
\end{align}
in terms of the normal vector to the boundary. This construction assumes that these components of the Weyl tensor are of holographic order, and in fact, this is the key argument to truncate Kounterterm charges in Einstein AdS~\cite{Olea:2005gb,Olea:2006vd} and in Einstein-Gauss-Bonnet AdS gravity~\cite{Kofinas:2006hr} to the linear order in the AdS curvature~\cite{Jatkar:2014npa, Jatkar:2015ffa}. However, note that the FG expansion of the electric part of the Weyl tensor 
\begin{align}
    W^{iz}_{jz}&=\frac{z^4}{\left(d-4\right) \ell_{\rm eff}^2} \Bigg\{\mathcal{B}^{i}_{\left(0\right)j} + \frac{\alpha }{\left(d-1\right)\Delta^{(1)} \ell^2_{\rm eff}}\bigg[6\left(d-2\right)(\mathcal{W}^2_{(0)})^i_j \notag \\ &-\frac{4d^2-d-12}{d}\delta^i_j|\mathcal{W}_{(0)}|^2 \bigg]\Bigg\}+\mathcal{O}(z^6)\,,
\end{align}
implies that the Ashtekar-Magnon-Das (AMD) charges are finite only for AAdS spaces with a conformally flat boundary in the non-degenerate case.

Along this line, AMD-like formulas were derived for Lovelock AdS in Refs.~\cite{Arenas-Henriquez:2017xnr,Arenas-Henriquez:2019rph}. They are proportional to the degeneracy condition  
\begin{equation}
\Delta^{(1)}=\sum_{p=1}^{N}\frac{(-1)^{p+1}(d-2)!p\alpha_p }{(d-2p)!\ell_{\rm eff}^{2(p-1)}},
\end{equation}
given in terms of the $\alpha_p$ couplings (each one corresponding to the Lovelock scalar of order~$p$), and the maximal number of curvatures $N$. Due to the fact any theory of the Lovelock type is asymptotically Einstein (with a modified Newton's constant), it is clear that the electric part of the Weyl tensor will turn Conformal Mass divergent when evaluated in non-ACF spacetimes~\cite{Dotti:2005rc,Ray:2015ava}. Dealing with this sort of infrared divergences will be the topic of future investigation. However, one may anticipate that, in even spacetime dimensions, additional boundary terms prescribed by higher-dimensional Conformal Invariants may play an essential role in producing renormalized action and charges. In doing so, one may conclude that the holographic stress tensor for an arbitrary Lovelock-AdS gravity is given~by
\begin{equation}
\langle T_{ij} \rangle = -\frac{d\Delta^{(1)}}{16\pi G\,\ell_{\rm eff}} \, g_{(d)ij},
\end{equation}
for odd boundary dimension $d$, as this coefficient is the only object with the suitable conformal weight. The presence of the degeneracy condition not only impedes the linearization of the field equations around an AdS background, but poses an obstruction to a standard holographic description of the theory.

\begin{acknowledgments}
We thank Loreto Osorio for participating in the early stages of this work. IJA gratefully acknowledges support from the Simons Center for Geometry and Physics, Stony Brook University, at which some of the research for this paper was performed. RO would like to thank the organizers of the Workshop “Conformal higher spins, twistors and boundary calculus” at U. Mons, for hospitality and discussions on Conformal Renormalization. This work is partially funded by Agencia Nacional de Investigación y Desarrollo (ANID) through the FONDECYT grants No. 1240043, 1240048, 1230112, 1230492, 1231779, 1240955, 1210500, 11230419, 11240059,  1231133 and 3230222.
\end{acknowledgments}

\appendix

\section{Appendix: Radial foliation of the spacetime} \label{GC}

For the radial decomposition of the tensorial objects, the metric in Gaussian coordinates is expressed as
\begin{equation}
ds^2 = N^2 \left(z\right) \diff{z^2} + h_{ij} \left(z, x^i\right) \diff{x^i} \diff{x^j} \,,
\end{equation}
where $N\left(z\right)$ is the lapse function. The normal to the constant $z$ surfaces takes the form
\begin{equation}
    n_{\mu} = -N \delta^z_{\mu} \,.
\end{equation}
On the other hand, the extrinsic curvature is defined as
\begin{equation}
K_{ij} = - \frac{1}{2N} \partial_z h_{ij} \,.
\end{equation}
The independent components of the Christoffel symbols are written as
\begin{equation}
\Gamma^z_{ij} = \frac{1}{N} K_{ij} \,, \quad \Gamma^{i}_{zj} = -N K_j^i \,, \quad \Gamma^i_{jk}\left(g\right) = \Gamma^i_{jk} \left(h\right) \,.
\end{equation}
The above expressions lead to the Gauss-Codazzi relation for a radial foliation of the spacetime
\begin{align}
R^{ij}_{km} &=\mathcal{R}^{ij}_{km} \left(h\right)- \left(K^i_k K^j_m -K^i_m K^j_k\right) \,, \\
R^{iz}_{mz}&=-\frac{1}{N} \partial_z K^i_m - K^i_s K^s_m \,, \\
R^{iz}_{jm}&=-\frac{1}{N} \left(\mathcal{D}_j K^i_m - \mathcal{D}_m K^i_j\right) \,, \\
R&=\mathcal{R}\left(h\right) - \left(K^2 + K^i_j K^j_i\right) -\frac{2}{N} \partial_z K \,.
\end{align}

\bibliographystyle{JHEP}
\bibliography{EGBbiblio.bib}
\end{document}